\documentclass[english,aps,twocolumn]{revtex4}
\usepackage[T1]{fontenc}
\usepackage[latin1]{inputenc}
\usepackage{array}
\usepackage{subfigure}
\usepackage{float}
\usepackage{color}
\usepackage{graphicx}
\usepackage{amssymb}

\makeatletter

\providecommand{\tabularnewline}{\\}

\newcommand\munit{$\mathrm{GeV}/c^2$}

\usepackage{babel}
\makeatother
\begin{document}


\title{Prospects of Searches for Neutral, Long-Lived Particles which Decay
to Photons using Timing at CDF}

\author{David Toback}

\email{toback@fnal.gov}

\affiliation{Texas A\&M University, College Station, Texas 77843-4242}

\author{Peter Wagner}

\email{wagnp@fnal.gov}

\affiliation{Texas A\&M University, College Station, Texas 77843-4242}

\begin{abstract}
We present the prospects of searches for neutral, long-lived particles
which decay to photons using their time of arrival measured with a
newly installed timing system on the electromagnetic calorimeter (EMTiming)
of the Collider Detector at Fermilab (CDF). A Monte Carlo simulation
shows that EMTiming can provide separation between decay photons from particles with both a long lifetime and a low boost, and prompt photons from Standard Model backgrounds. Using
a gauge mediated supersymmetry breaking (GMSB) $\tilde{\chi}_{1}^{0}\rightarrow\gamma\tilde{G}$
model we estimate a quasi-model-independent sensitivity using only
direct neutralino pair production, and also estimate the expected
95\%~confidence level exclusion regions for all superpartner production as a function
of the neutralino mass and lifetime. We find that a combination of
single photon and diphoton analyses should allow the Tevatron in run~II
to easily extend the exclusion regions from LEP~II at high neutralino
masses and lifetimes, and cover much, if not all, of the theoretically
favored $m_{\tilde{G}}<1$~$\mathrm{keV}/c^2$ parameter space for
neutralino masses less than 150~\munit.
\end{abstract}
\maketitle

\section{introduction}

The electromagnetic (EM) calorimeter at the Collider Detector at Fermilab
(CDF) \cite{key-25} has recently been equipped with a new nanosecond-resolution
timing system, EMTiming~\cite{key-24}, to measure the arrival time
of energy deposited (e.g. from photons). While
it was initially designed to reject cosmics and accelerator backgrounds
\cite{key-10}, we investigate the possibility of using it to search
for neutral particles~\cite{key-29} with a lifetime of the order
of a nanosecond which decay in flight to photons. An example of
a theory which would produce these particles is the gauge mediated supersymmetry
breaking (GMSB) model \cite{key-2} with a neutralino, $\tilde{\chi}_{1}^{0}$, as the
next-to-lightest supersymmetric particle (NLSP) and a light gravitino,
$\tilde{G}$, as the LSP. \textcolor{black}{In this scenario the neutralino
decays preferably ($\sim$100\%) as} \textcolor{black}{\small $\tilde{\chi}_{1}^{0}\rightarrow\gamma\tilde{G}$}
with a macroscopic lifetime for much of the
GMSB parameter space.

\begin{figure}
\begin{center}\includegraphics[%
  width=8.6cm,
  keepaspectratio]{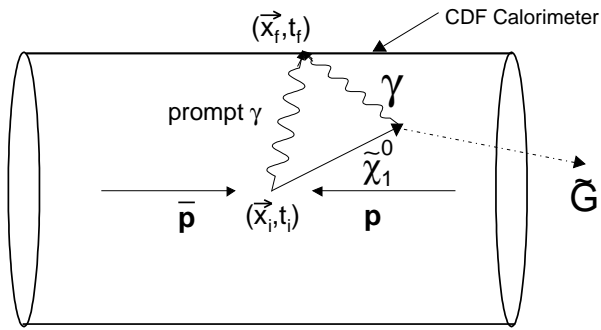}\end{center}

\caption{\label{cap:Explanation-of-Delta}A schematic diagram of a long-lived neutralino decaying to a photon and
a gravitino in the CDF detector. The neutralino emanates from the collision at $(\vec{x}_{i},t_{i})$
and after a time $\tau$ it decays. While the gravitino leaves the
detector the photon travels to the detector wall and deposits energy
in the EM calorimeter where its final location $\vec{x}_{f}$ and
arrival time $t_{f}$ can be measured. A prompt photon would 
travel directly from $\vec{x}_{i}$ to $\vec{x}_{f}$ which
can both be measured. The difference between the actual time the neutralino/photon
needs, $\Delta t=t_{f}-t_{i}$, and the time a prompt photon would
need, $\frac{|\vec{x}_{f}-\vec{x}_{i}|}{c}$, is defined as
$\Delta s$. The SM typically produces prompt photons which have $\Delta s$~=~0~ns,
whereas photons from delayed decays from SUSY have \mbox{$\Delta s>0$~ns,}
assuming a perfect measurement.}
\end{figure}

We begin with a study of the properties of events where timing can be used to separate between
decay photons from long-lived particles, and photons produced promptly at the collision. A suitable variable to describe this distinction
is the measured difference between the time after which the photon
arrives at the face of the detector, and the time a prompt photon
would virtually need to reach the same final position. This time difference
for a prompt photon, from Standard Model (SM) sources, is exactly
0 but is always greater than~0 for photons from delayed decays, as in GMSB/Supersymmetry,
if we neglect the measurement resolution. We call this difference
$\Delta s$:

\textcolor{black}{\begin{equation}
\Delta s\equiv(t_{f}-t_{i})-\frac{|\vec{x}_{f}-\vec{x}_{i}|}{c}\label{eq:DeltaseqDeltatxfx0}\end{equation}
where $t_{f}-t_{i}$ is the time between the collision and the arrival
time of the photon at the face of the detector, and} $|\vec{x}_{f}-\vec{x}_{i}|$ \textcolor{black}{is
the distance between the final position of the photon and the collision
point. The situation is visualized in Fig.}~\ref{cap:Explanation-of-Delta}\textcolor{black}{.
All four variables can be measured by the CDF detector~\cite{key-30}
and give a system resolution of $\sigma_{\mathrm{EMTiming}}$ $\sim1.0$~ns~\cite{key-26}. }

An important note in our analysis is that photons from long-lived
particles will usually not arrive at the face of the EM calorimeter
at the usual 90 degree incident angle. This could have serious implications
for photon identification. For the purposes of this study we assume
that this issue can be addressed without significant changes to
the identification efficiency. We further assume that the additional handles such
as EMTiming and timing in the hadronic calorimeters provide the necessary
robustness needed to convince ourselves that photons which might not
pass ordinary selection requirements are indeed from our signal source
as opposed to sources which could produce fake photons and missing
transverse energy, $E_{T}\!\!\!\!\!\!\!/\,\,\,$, like cosmics.

We estimate our sensitivity to two different types of new particle production using GMSB models. 
As a quasi-model-independent sensitivity estimate to generic long-lived particles we simulate direct neutralino pair production and decay, 
and examine the dependency as a function of both neutralino
mass, $m_{\tilde{\chi}}$, and lifetime, $\tau_{\tilde{\chi}}$. For a ``full'' GMSB model sensitivity, which
means including all relevant GMSB subprocesses such that the neutralinos
are part of cascades from gauginos and squarks, we allow all SUSY particle production and decay, 
and again vary the mass and lifetime variables. 
To choose analysis final states for both we consider three
issues: 1) with neutralino lifetimes longer than a  nanosecond it
is possible that one or both of the neutralinos leave the detector
before they decay, 2) with gravitinos or the neutralino leaving
the detector $E_{T}\!\!\!\!\!\!\!/\,\,\,$ should also help separate
signal from SM backgrounds and 3) to ensure
that our predictions are as reliable as possible we want to use 
the data selection requirements and 
background predictions from previously published papers by CDF~\cite{key-25} and D\O~\cite{key-41}.
In the 1992-1995 collider run (run~I) of the Tevatron three types of analyses match these criteria: CDF and D\O\ results in $\gamma\gamma$~+~$E_{T}\!\!\!\!\!\!\!/\,\,\,$~\cite{key-10,key-55},
exclusive $\gamma$~+~$E_{T}\!\!\!\!\!\!\!/\,\,\,$ ($\gamma$~+~$E_{T}\!\!\!\!\!\!\!/\,\,\,$~+~0~jets)
from CDF~\cite{key-7}  and $\gamma$~+~$E_{T}\!\!\!\!\!\!\!/\,\,\,$~+~jets from D\O~\cite{key-8}.
Since there are no jets at the parton level in direct neutralino pair
production, in this case we consider analyses with final states $\gamma\gamma$~+~$E_{T}\!\!\!\!\!\!\!/\,\,\,$
and $\gamma$~+~$E_{T}\!\!\!\!\!\!\!/\,\,\,$~+~0~jets. To estimate
the sensitivity for full GMSB neutralino production we consider both $\gamma\gamma$~+~$E_{T}\!\!\!\!\!\!\!/\,\,\,$
and $\gamma$~+~$E_{T}\!\!\!\!\!\!\!/\,\,\,$~+~jets analyses. 

For both direct neutralino pair production and full GMSB production we quantify the sensitivity 
for 2~fb$^{-1}$ in run~II using the
expected 95\% confidence level (C.L.) cross section upper limits. Results for both with and without the EMTiming system, using kinematics cuts only, illustrate the 
contribution to the final sensitivity from kinematic and timing information considerations~\cite{key-39}. Finally, we compare the final mass and 
lifetime exclusion regions for a GMSB scenario to direct and indirect searches from the ALEPH experiment~\cite{key-3} at LEP~II 
and the theoretically favored parameter space from cosmological model restrictions of 
$m_{\tilde{G}}<1$~$\mathrm{keV}/c^2$~\cite{key-27}.

\section{Kinematic properties of events with long-lived particles which decay
to photons\label{sub:Introkinprop}\label{sub:Kinematics-Results}}

While the final sensitivity studies use both a full physics
generation and a detector simulation of the geometry and timing resolution,
we begin with a study of the kinematic properties of events which
yield large $\Delta s$ measurements using a ``toy Monte Carlo.'' For
now the CDF detector is assumed to be a cylinder, with length 3.5~m
and radius 1.7~m, instrumented with time and position detectors of
perfect resolution for both the collision point and where the photon
hits the face of the detector. Neutral particles, which we will refer
to as neutralinos, are simulated as emanating isotropically from the
center of the detector and emit a photon isotropically after a lifetime
$\tau_{\tilde{\chi}}$ in their rest frame. For a promptly decaying
photon the minimum time corresponds to the nearest distance to the
detector face: $\frac{1.7\mathrm{m}}{c}$~=~5.6~ns; the maximum
to the largest distance $\sqrt{(1.7\mathrm{m})^{2}+(\frac{3.5\mathrm{m}}{2})^{2}}/c$~=~8.1~ns.
For pedagogic reasons, neutralinos are simulated with a flat momentum
and lifetime distribution, i.e. independently any lifetime and momentum
have equal probability. We note that to be conservative here, as later in the paper, only those neutralinos that decay
before the face of the detector are considered to have produced a
photon. 

\begin{figure}
\begin{center}\includegraphics[%
  height=8.6cm,
  keepaspectratio]{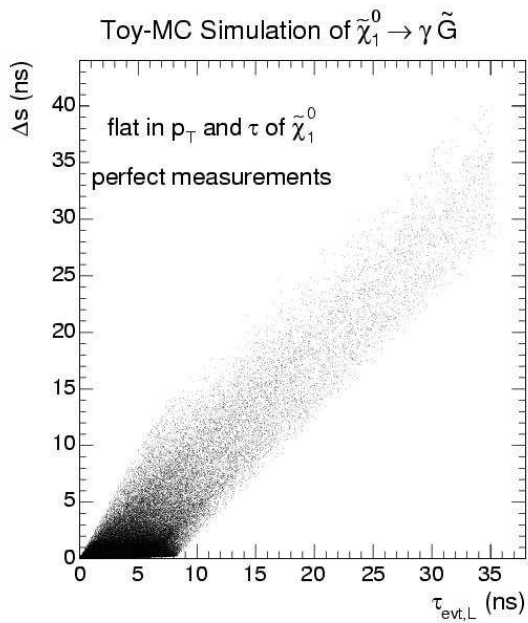}\end{center}

\caption{\textcolor{black}{\label{cap:Delta-s-(ns)}The $\Delta s$ distribution
as a function of the event lifetime in the neutralino lab frame for
a toy Monte Carlo simulation. In general, $\Delta s$ is proportional
to $\tau_{\mathrm{evt,L}}$. At large $\tau_{\mathrm{evt,L}}$ most
of the neutralinos leave the detector and are not shown
here. The spread perpendicular to $\Delta s\sim\tau_{\mathrm{evt,L}}$
originates in variations of the neutralino momentum as well as in
variations in the travel time of the photon due to detector geometry.
Essentially, events with large $\Delta s$ require a neutralino with
a long lifetime.}}
\end{figure}

Figure~\ref{cap:Delta-s-(ns)} shows the measured \textcolor{black}{$\Delta s$
versus the event lifetime of the $\tilde{\chi}_{1}^{0}$ in the lab
frame, $\tau_{\mathrm{evt,L}}$.}
\textcolor{black}{For
$\Delta s\gtrsim10$~ns there is a roughly linear relation between
$\Delta s$ and $\tau_{\mathrm{evt,L}}$. For a fixed $\tau_{\mathrm{evt,L}}$
the maximum $\Delta s$ (upper bound) occurs when the neutralino travels
to the farthest corner of the detector and then emits a photon backward
to the opposite corner. Analogously we get a minimum $\Delta s$ (lower
bound) if the neutralino travels with high momentum to the nearest
part of the detector and emits a photon forward. The latter would
look like a usual prompt photon event except for the difference in
velocity between the neutralino and the photon. If the event lifetime
is greater than the maximum time a prompt photon would need to travel
to the detector then $\Delta s$ is restricted from below and $\Delta s$~>~0~ns
(given that the neutralino decays inside the detector). Thus, the
spread mainly comes from detector geometry but with the neutralino
momentum also contributing to the width. }

\begin{figure}
\begin{center}\includegraphics[%
  width=8.6cm,
  keepaspectratio]{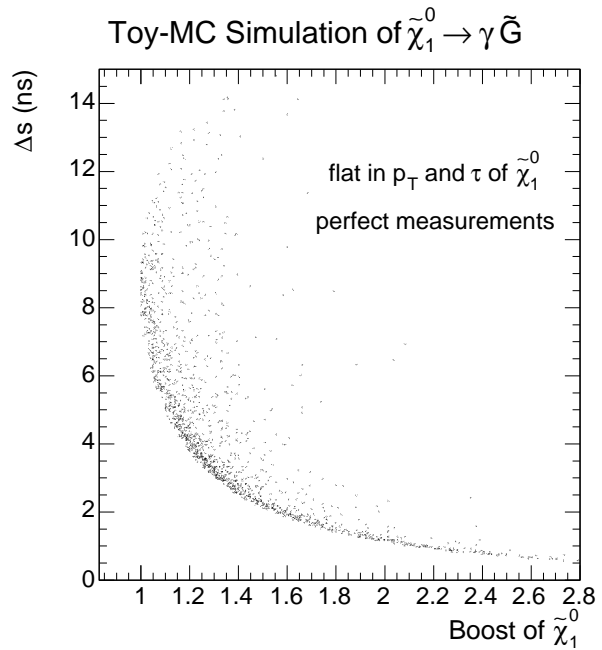}\end{center}

\caption{\textcolor{black}{\label{cap2:Delta-s-(ns)}The $\Delta s$ distribution
as a function of the boost of the neutralino for a lifetime
{}``slice'' of 8.5~ns~$\leq\tau_{\mathrm{evt,L}}\leq$~9.0~ns.
In the region 1.0~<~boost~<~1.5 neutralinos remain in the
detector and can produce a large $\Delta s$. Neutralinos with high boost, that is high $p_{T}$,
are more likely to leave the detector or, if they don't, produce low $\Delta s$.
Thus, events with the largest $\Delta s$ are produced by neutralinos with large
lifetimes and low boosts.}}
\end{figure}

\textcolor{black}{Figure~\ref{cap2:Delta-s-(ns)} shows $\Delta s$
versus the neutralino boost for the lifetime slice 8.5~ns~$\le\tau_{\mathrm{evt,L}}\leq$~9.0~ns.
A low boost (between 1.0 and 1.5) allows large $\Delta s$ since
neutralinos can have a larger lifetime without leaving the detector.
Neutralinos with high boost are more likely to leave the detector,
and even if they do not and their photon is detected, it has low $\Delta s$
(0~ns~$\lesssim\Delta s\lesssim$~2~ns). Thus, events with large
$\Delta s$ are produced by neutralinos with long lifetimes and low
boost.}

\begin{figure}
\begin{center}\includegraphics[%
  width=8.6cm,
  keepaspectratio]{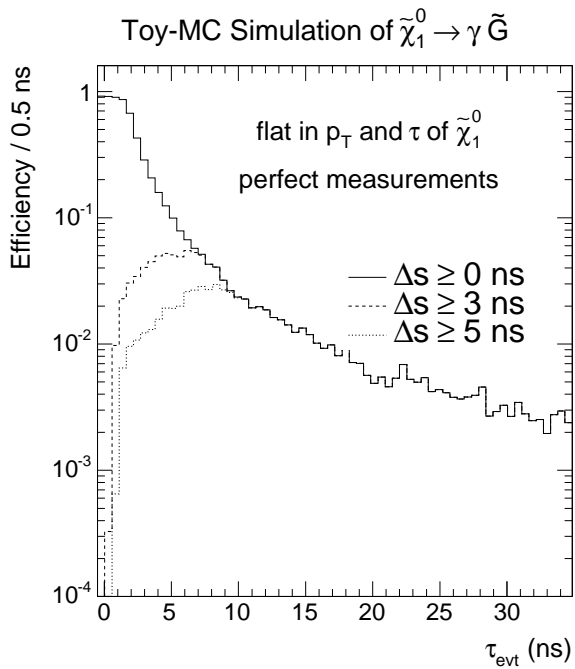}\end{center}

\caption{\textcolor{black}{\label{cap:Efficiency-vs.-tau}The efficiency as
a function of the event lifetime, $\tau_{\mathrm{evt}}$, of the neutralino.
We distinguish between events in which the neutralino remains in the
detector, and events with photons of medium and large $\Delta s$.
The efficiency is 100\% for prompt} decays (a small difference shows up
as a binning effect) for a photon to be identified, but \textcolor{black}{only
a very small efficiency for events with low $\tau_{\mathrm{evt}}$
at large $\Delta s$. At large $\tau_{\mathrm{evt}}$ only few events
stay in the detector, however if a neutralino is long-lived and stays
in the detector, it has large $\Delta s$. We note that the true efficiency
shape depends on the production mechanism i.e. the neutralino $p_{T}$
distribution. }}
\end{figure}

Next we consider the efficiency for neutralinos to remain in the detector and/or 
produce a photon with large $\Delta s$. 
Figure~\ref{cap:Efficiency-vs.-tau} shows the efficiency, the fraction of all generated events that
produce photons which pass a given $\Delta s$ restriction, as a function of the event lifetime, $\tau_{\mathrm{evt}}$, for $\Delta s\geq$~0 
(neutralino stays in the detector),~3~ns and 5~ns for the same production distribution. \textcolor{black}{While 
these results change for a more realistic $p_T$ spectrum, the qualitative features are instructive. In the limit of} 
$\tau_{\mathrm{evt}}$\textcolor{black}{~=~0~ns
and $\Delta s\geq0$} the efficiency is 100\% and the efficiency decreases
with higher event lifetime, since the neutralinos are more likely
to leave the detector. When one applies a $\Delta s$ cut however,
there is no efficiency for events that contain neutralinos with a
low event lifetime ($\tau_{\mathrm{evt}}\lesssim$~2~ns). \textcolor{black}{For
any $\Delta s>0$ requirement the efficiency goes to 0\% at} $\tau_{\mathrm{evt}}$\textcolor{black}{~=~0~ns,
since all photons would have $\Delta s=0$. A higher $\Delta s$ cut
gradually suppresses events with a neutralino lifetime of about $\tau_{\mathrm{evt}}\lesssim2\cdot\Delta s$,
whereas it does not suppress any events with a high lifetime.} So,
if an event contains a neutralino with a long lifetime and which decays
in the detector, the decay photon always has high \textcolor{black}{$\Delta s$. }

\section{neutralino pair production as a measure of quasi-model-independent
sensitivity}

\subsection{Analysis Methods and their Efficiency as a Function of Neutralino
Mass and Lifetime\label{sub:Analysis-Methods-and}}

Here we estimate our sensitivity to neutral, long-lived particles
which decay to photons in as model-independent a manner as possible.
We do so by considering a GMSB model~\cite{key-23} which we restrict to direct neutralino pair
production and decay: $p\bar{p}\rightarrow\tilde{\chi}_{1}^{0}\tilde{\chi}_{1}^{0}\rightarrow\gamma\tilde{G}\gamma\tilde{G}$.
We use the \small{PYTHIA}~\cite{key-5} event generator, with \small{ISAJET}~\cite{key-56} to generate the SUSY masses, and PGS
with the parameter file for the CDF detector~\cite{key-6} as a simple
detector simulation, modified for the use of timing information. \textcolor{black}{We
accept photons with a rapidity $|\eta|$~$\le$~2.1 and a transverse
energy $E_{T}$~$\geq$~12~GeV according to the CDF/EMTiming  fiducial region
and trigger~\cite{key-25,key-24}.} We first look at the efficiency of the
timing system with infinite resolution as a function of neutralino
mass and lifetime for different $\Delta s$ restrictions. Then we
discuss background estimations, take into account a timing resolution
of 1.0~ns and find the sensitivities for the model predictions for
both the single and diphoton analysis.

\begin{figure}
\begin{center}{\includegraphics[%
  height=8.6cm,
  keepaspectratio]{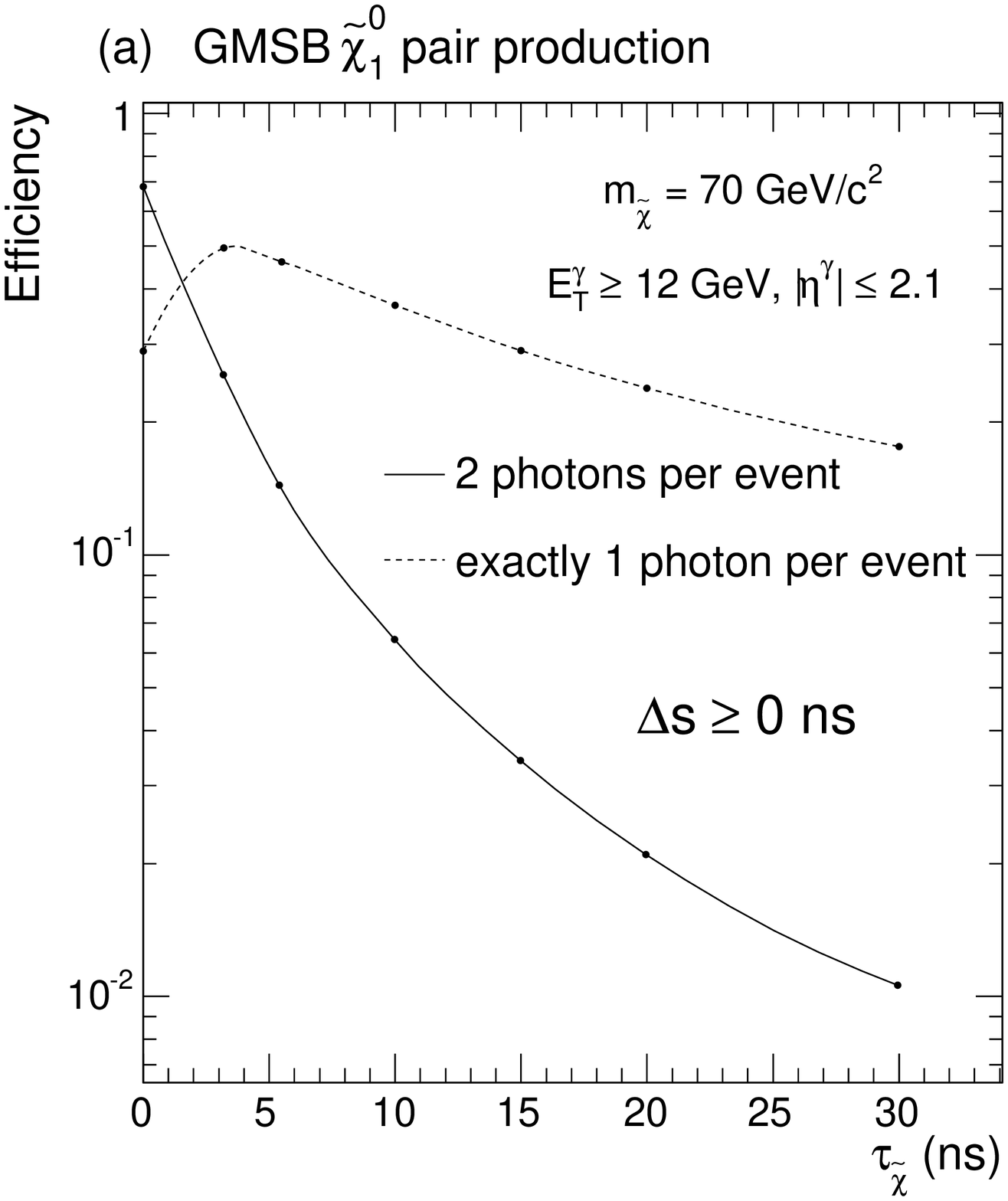}}
{\includegraphics[%
  height=8.6cm,
  keepaspectratio]{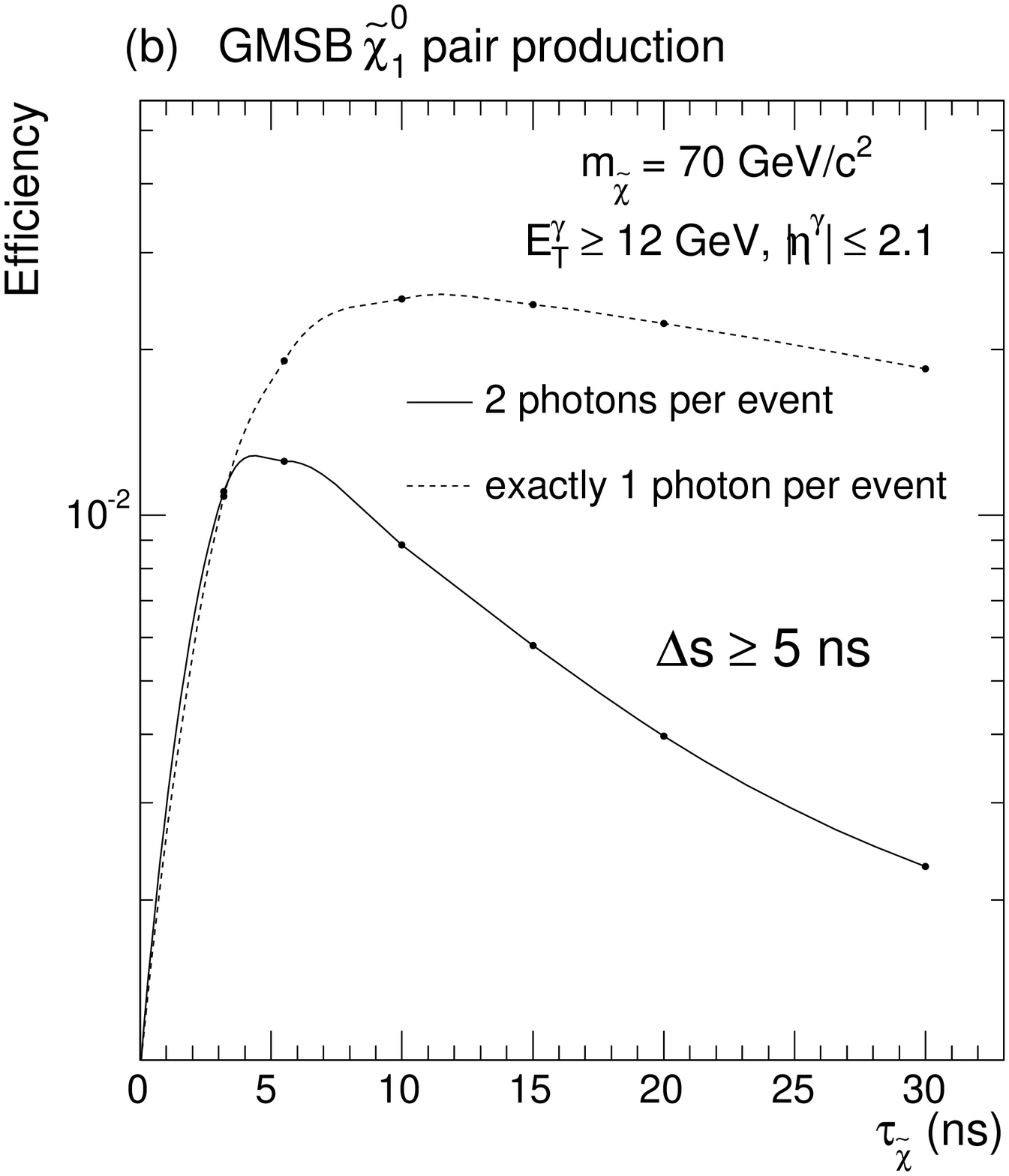}}\end{center}

\caption{\textcolor{black}{\label{cap2:Efficiency-vs.-tau}The efficiency
for events to pass the various $\Delta s$-cuts (assuming perfect
measurements) as a function of the neutralino lifetime at constant
mass ($m_{\tilde{\chi}}$=~70~}\munit\textcolor{black}{), separated into
single and diphoton events at $\Delta s$ $\ge$ 0~ns and 5~ns.
For any $\Delta s>0$ the efficiency is zero at $\tau_{\tilde{\chi}}$~=~0~ns,
since all photons would have $\Delta s=0$. For high $\tau$ neutralinos
have a higher probability to leave the detector. For any $\Delta s$
one can find an efficiency maximum at about 5-10~ns. Single photon
events are preferred towards higher $\Delta s$ requirements and/or higher lifetimes,
due to increasing probability for a photon to leave the detector.
Thus, we expect a $\gamma\gamma$~+~}$E_{T}\!\!\!\!\!\!\!/\,\,\,$
\textcolor{black}{to provide the best sensitivity for very low lifetimes, and
a $\gamma$~+~}$E_{T}\!\!\!\!\!\!\!/\,\,\,$ \textcolor{black}{analysis to be best
for higher lifetimes.}}
\end{figure}

Figure~\ref{cap2:Efficiency-vs.-tau} shows the efficiency \textcolor{black}{versus neutralino lifetime}
for a mass of~70~\munit~\cite{key-11}
\textcolor{black}{for events with $\Delta s$~$\ge$~0~ns
(photons from neutralinos remaining in the detector) and events with
a $\Delta s$~$\ge$~5~ns, separated into} single and diphoton
events\textcolor{black}{. We get
essentially the same shapes as in Fig.~\ref{cap:Efficiency-vs.-tau},
however the overall efficiency is less as we now consider the exponentially
distributed neutralino lifetime instead of an event lifetime.} For
all four distributions there is an efficiency maximum in the lifetime
region between 4 and 9~ns. At lower lifetimes the probability that
the neutralino stays in the detector is large enough that the diphoton
final state dominates. \textcolor{black}{For any $\Delta s>0$ requirement
the efficiency is zero at} $\tau_{\tilde{\chi}}$\textcolor{black}{~=~0~ns,
since all photons have $\Delta s=0$}. At a lifetime of about 3~ns,
independent of the \textcolor{black}{$\Delta s$ cut,} single photon
events become dominant. \textcolor{black}{At high lifetimes the efficiency
decreases rapidly for both analyses as most of the neutralinos leave
the detector. Hence, in order to have sensitivity in as much lifetime
range as possible, we consider both $\gamma$~+~}$E_{T}\!\!\!\!\!\!\!/\,\,\,$
\textcolor{black}{and $\gamma\gamma$~+~}$E_{T}\!\!\!\!\!\!\!/\,\,\,$
\textcolor{black}{analyses. }

\begin{figure}
\begin{center}{\includegraphics[%
  height=8.6cm,
  keepaspectratio]{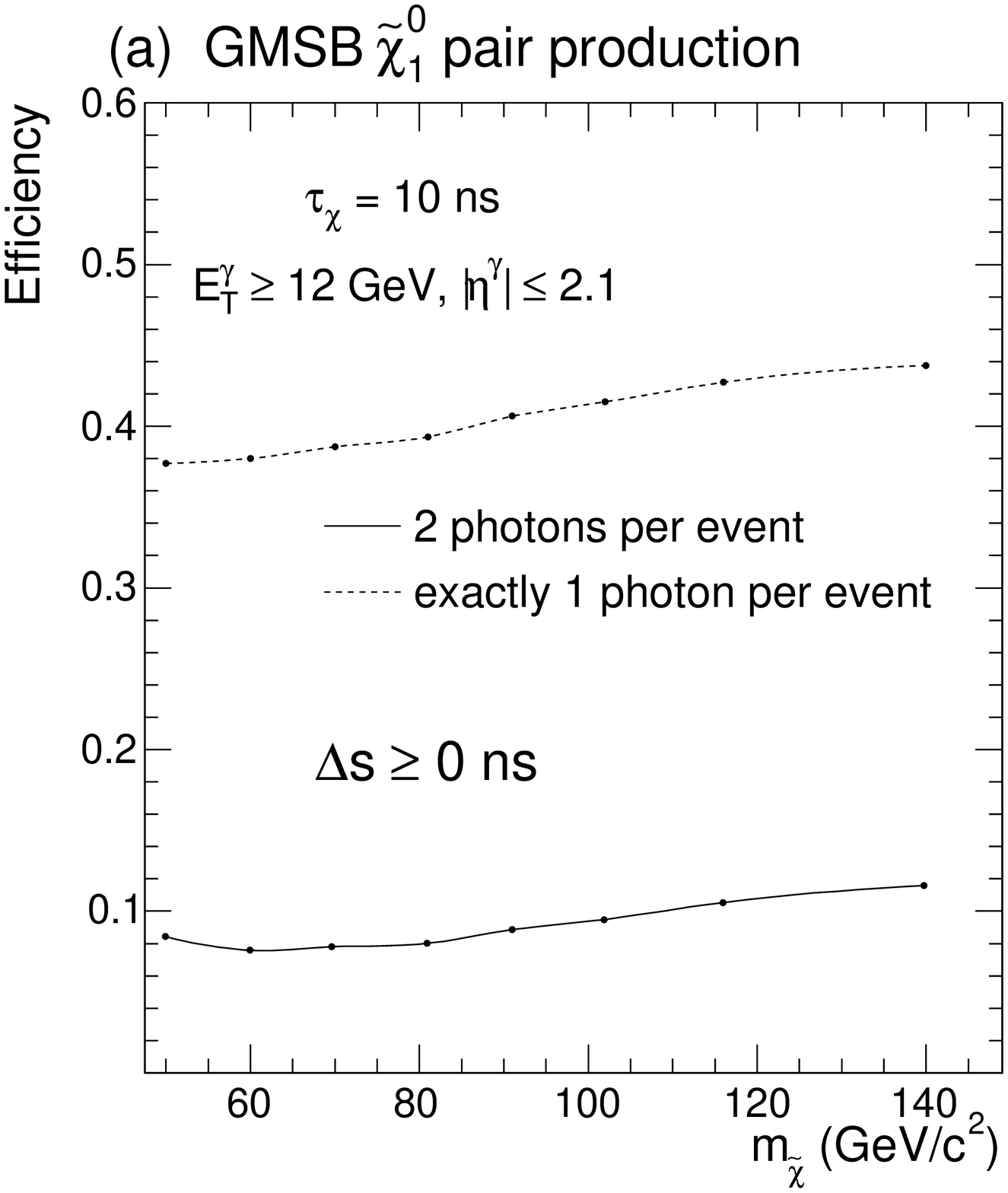}}
{\includegraphics[%
  height=8.6cm,
  keepaspectratio]{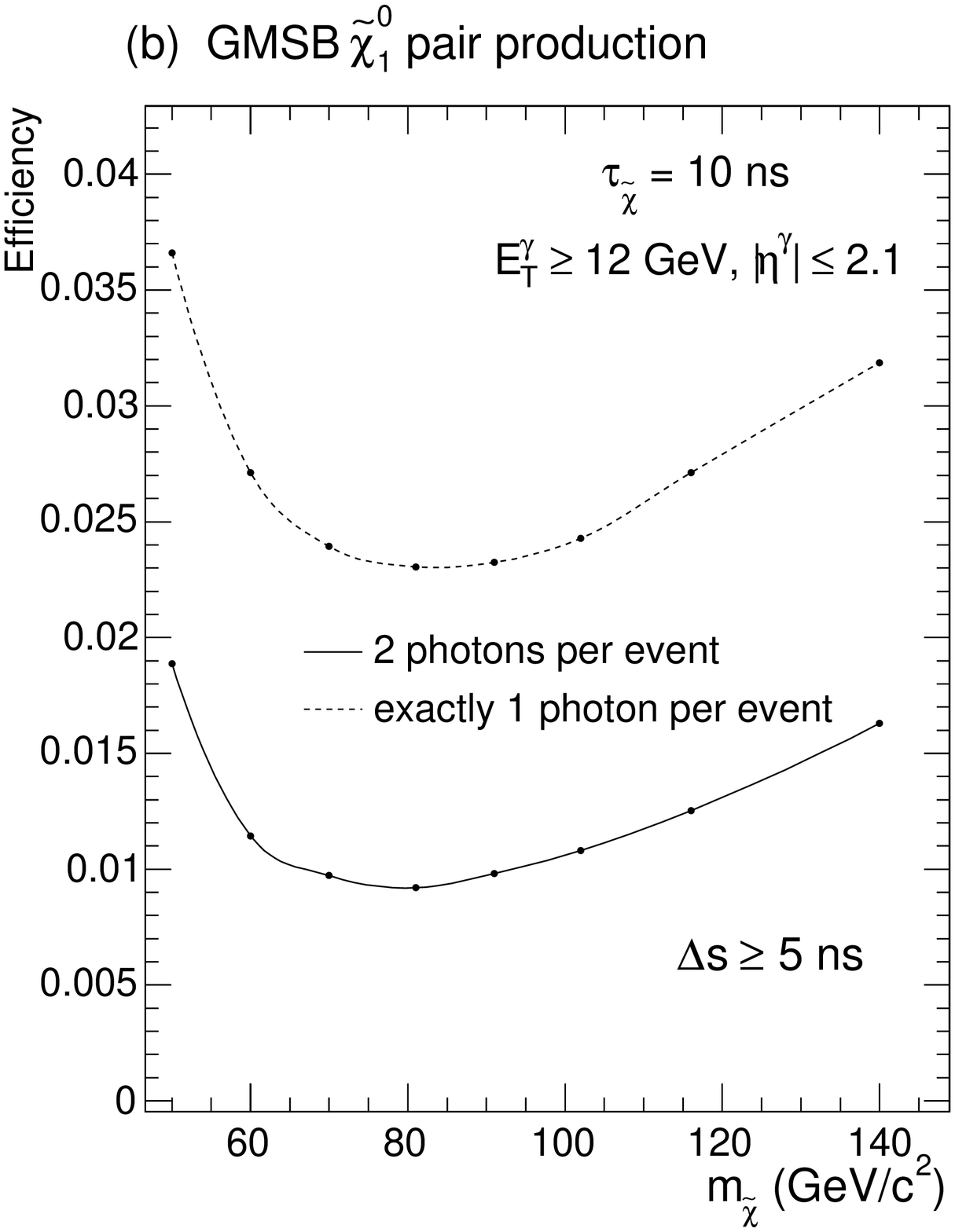}}\end{center}

\caption{\textcolor{black}{\label{cap:Efficiency-vs.-m_N1}The efficiency
as a function of the neutralino mass at a lifetime} $\tau_{\tilde{\chi}}$\textcolor{black}{~=~10~ns
for single and diphoton events at $\Delta s$ $\ge$~0~ns and 5~ns
(assuming perfect measurements). The ratio of single to diphoton
events is independent of the neutralino mass and is roughly constant
as a function of $\Delta s$. One can see a soft {}``dip'' in the
efficiency curve in a mass range of 40~}\munit$-$80~\munit\textcolor{black}{.
This effect is production dependent and due to a change in the $p_{T}$
distribution of the neutralinos (see Fig. \ref{cap:Neutralino-p_T}).}}
\end{figure}

\begin{figure}
\begin{center}\includegraphics[%
  height=8.5cm,
  keepaspectratio]{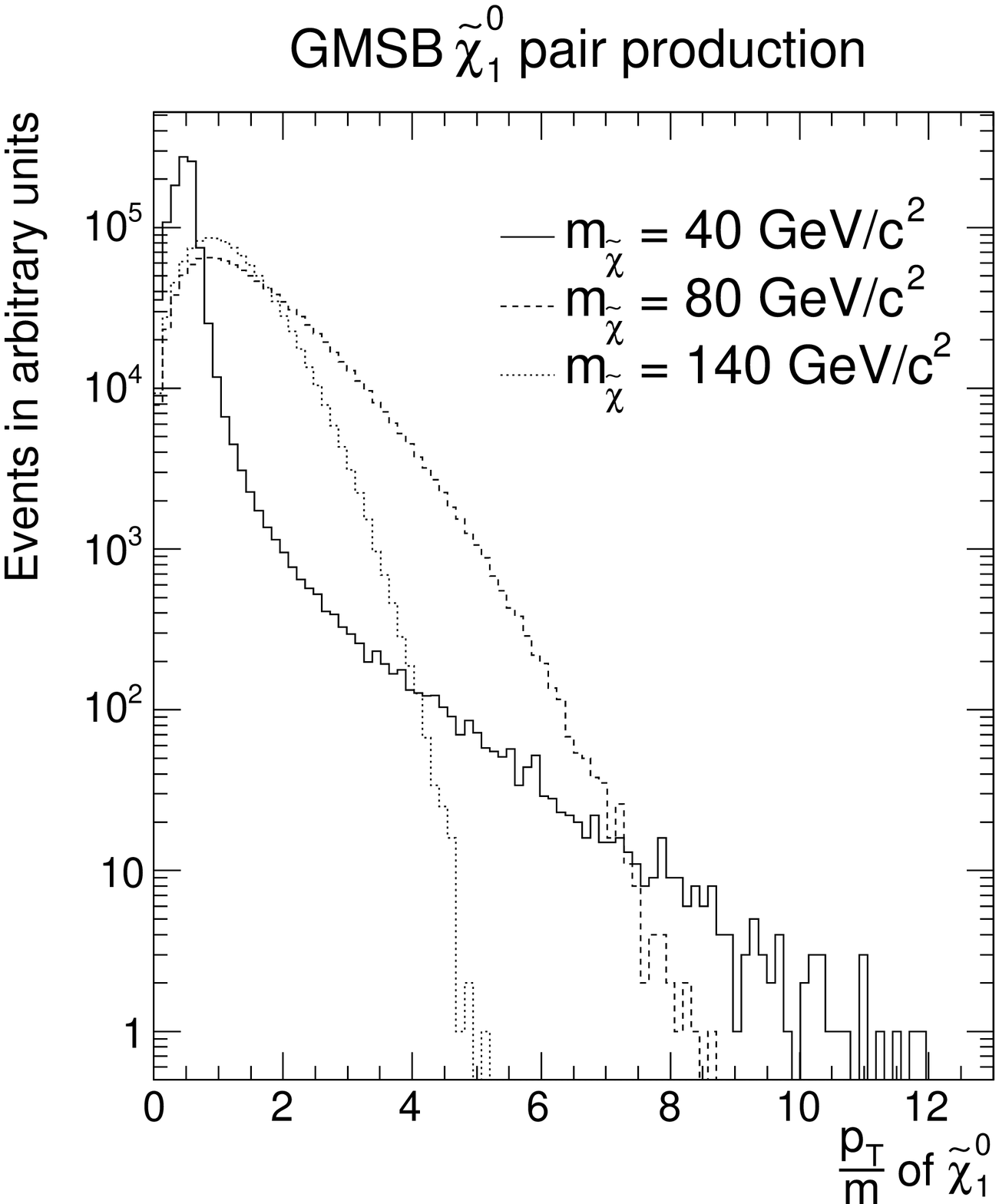} \end{center}

\caption{\label{cap:Neutralino-p_T} The neutralino $\frac{p_{T}}{m}$ distribution
for masses 40~$\mathrm{GeV}/c^2$, 80~\munit\ and 140~\munit\ for
neutralino pair production. For a mass of 80~\munit\
the maximum moves towards higher $\frac{p_{T}}{m}$ and the distribution
broadens compared to 40~\munit, yielding a greater fraction of high $p_{T}$ neutralinos
which either leave the detector or produce low $\Delta s$ photons, and thus a loss in efficiency. 
For higher masses the maximum remains
constant and the distribution narrows so the efficiency rises.}
\end{figure}

In contrast the timing efficiency is essentially constant as a function
of neutralino mass at a fixed lifetime. Figure~\ref{cap:Efficiency-vs.-m_N1}
shows the efficiencies at $\tau_{\tilde{\chi}}=10$~ns,
where the system has the highest efficiency and single photon
events dominate.
Note that the {}``dip'' in the efficiency can be explained by the
neutralino pair production mechanism: if the decay length is greater
than the distance to the detector wall, the neutralino will leave.
Since this is proportional to the ratio of the neutralino's transverse
momentum to its mass, $\frac{p_{T}}{m}$, (at constant lifetime),
the dip occurs from a change in the shape of the $\frac{p_{T}}{m}$
distribution of the neutralinos as shown in Fig.~\ref{cap:Neutralino-p_T}.
For a mass of 80~\munit\ the maximum moves towards higher $\frac{p_{T}}{m}$
and the distribution broadens compared to 40~\munit, yielding a
greater fraction
of high-$p_{T}$ neutralinos and hence a loss in efficiency. As the
mass gets higher the maximum remains the same and the distribution
narrows, which in turn leads to a gain in efficiency. Thus, the efficiency
is essentially independent of the neutralino mass, with slight variations
originating from the production mechanism, specifically the neutralino
momentum distribution.

\subsection{Backgrounds and Sensitivity to Neutralino Pair Production\label{sub:Sensitivity-to-N1pairs}}

We now estimate the sensitivity of our system in a quasi-model-independent
manner using the neutralino pair production introduced in the previous section, but taking
into account SM backgrounds, more realistic cuts and the timing resolution.
We consider separately single photon and diphoton events and present
our sensitivity as the expected 95\%~C.L. cross section upper limits for either case
assuming no signal in the data. We also compare our results with the results
of no EMTiming system available to estimate the effect of the EMTiming
system over a set of kinematics-only selection requirements. Throughout this section we use the relative
systematic uncertainties for luminosity, acceptance and background
rates given in Table~\ref{cap:Errors} and a $\Delta s$ resolution
of 1.0~ns. The expected cross section limits are calculated following
\cite{key-21} with the number of events observed {}``in the data''
fluctuating around an expected mean background rate according to Poisson
statistics. \textcolor{black}{The cross section limit is, for a certain
luminosity, a function of background events and signal acceptance,
where both in turn are functions of specified cuts (e.g. $\Delta s$
and} $E_{T}\!\!\!\!\!\!\!/\,\,\,$ \textcolor{black}{cuts in the} $\gamma\gamma$~+~$E_{T}\!\!\!\!\!\!\!/\,\,\,$
\textcolor{black}{case). By varying the cuts we find a signal acceptance
and number of background events that, after smearing by systematic
errors, minimizes the cross section limit. }

\begin{table}

\caption{\textcolor{black}{\label{cap:Errors}The systematic uncertainties, 
estimated based on Refs. \cite{key-8, key-10},
for luminosity, acceptance and number of background events for use in all 
analyses in estimating cross section limits.}}

\begin{center}\textcolor{black}{}\begin{tabular}{lcp{8.6cm}}
\hline
\hline 
Factor&
Syst. Uncertainty\tabularnewline
\hline
Luminosity&
5\% \tabularnewline
Acceptance&
10\% \tabularnewline
Number of background events&
30\% \tabularnewline
\hline
\hline
\end{tabular}\end{center}
\end{table}

\subsubsection{\label{sub:gammagamma-MET}$\gamma\gamma$ + $E_{T}\!\!\!\!\!\!\!/\,\,\,$}

\begin{figure*}
\begin{center}{\includegraphics[%
  width=5cm,
  keepaspectratio]{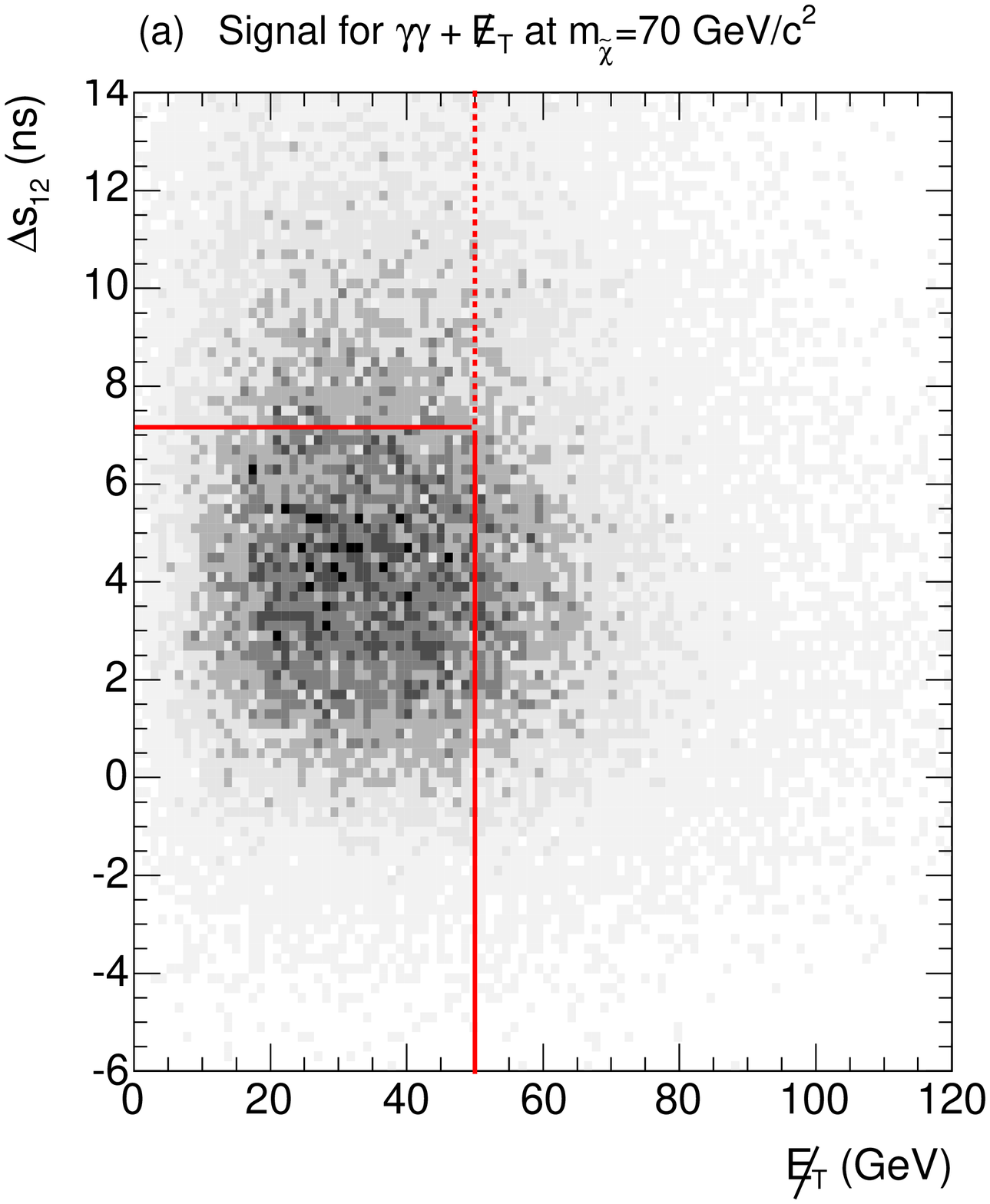}}
{\includegraphics[%
  width=5cm,
  keepaspectratio]{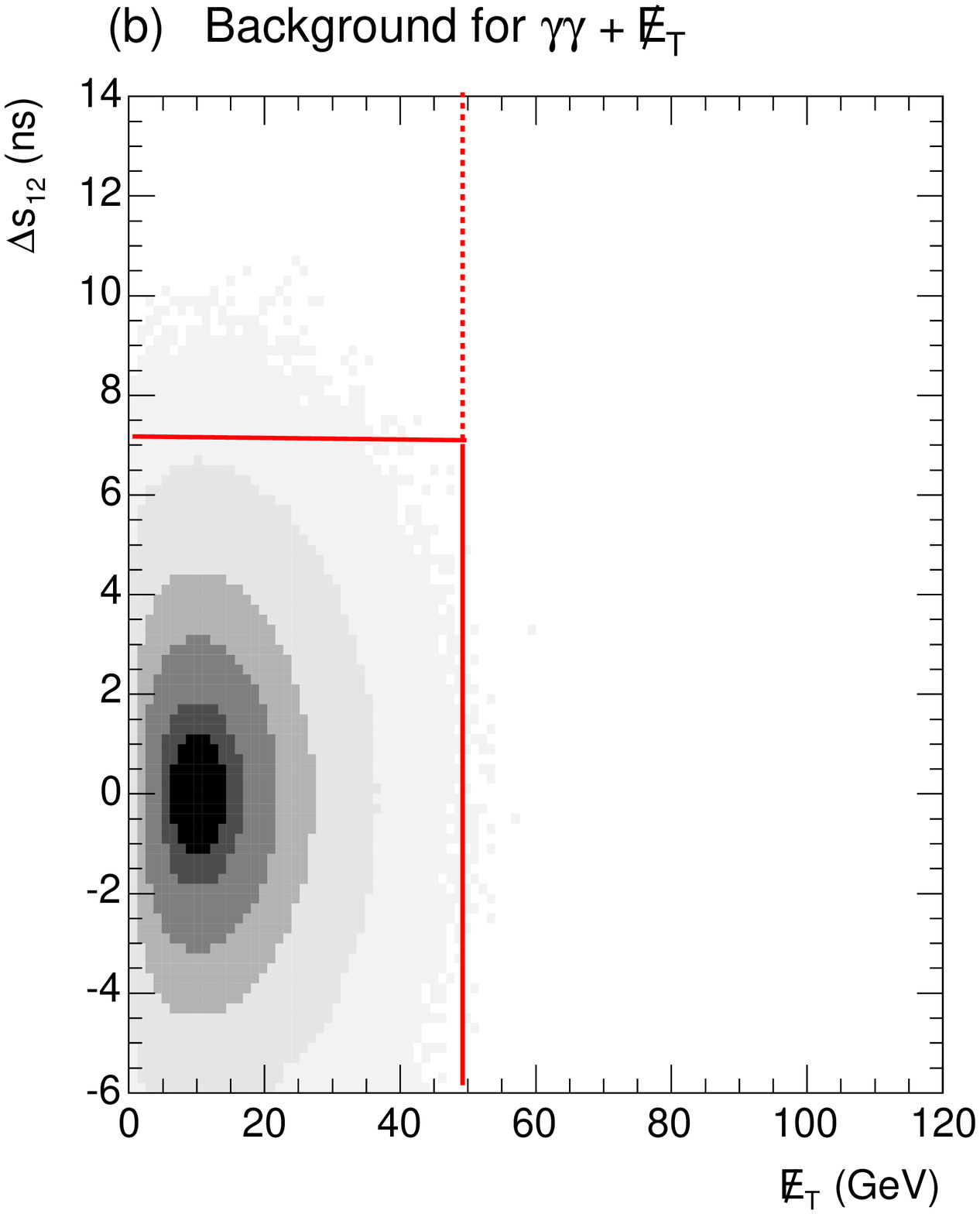}}
{\includegraphics[%
  width=5cm]{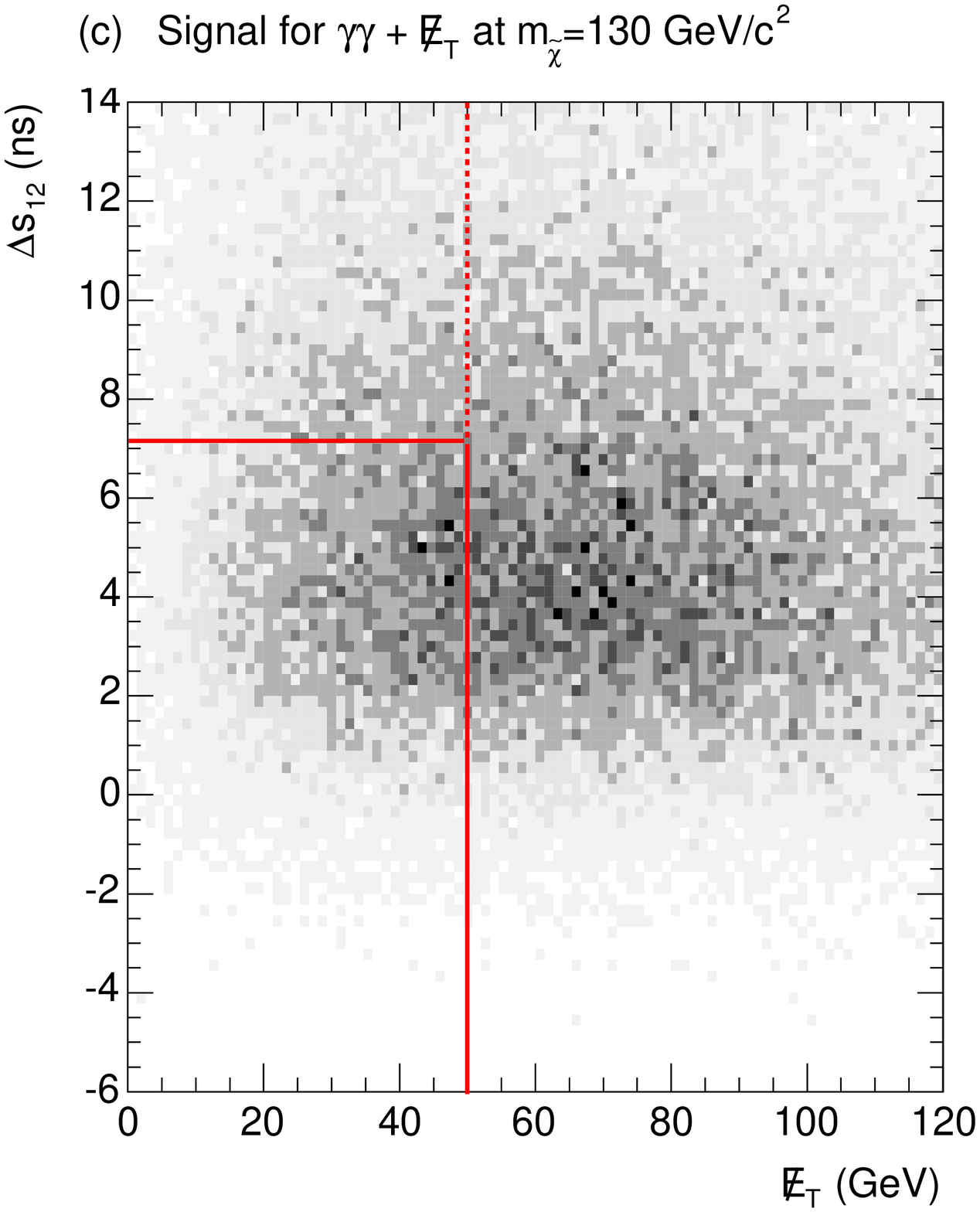}}
\end{center}

\caption{\label{cap:SigBkgdistribution}The distribution of $\Delta s_{12}$
and $E_{T}\!\!\!\!\!\!\!/\,\,\,$ for signal and background in the $\gamma\gamma$~+~$E_{T}\!\!\!\!\!\!\!/\,\,\,$
analysis. The distributions are (a) from direct neutralino pair production, with $m_{\tilde{\chi}}=70$~\munit\
and $\tau_{\tilde{\chi}}=10$~ns, (b) QCD background and (c) direct
neutralino pair production, with $m_{\tilde{\chi}}=130$~\munit\ and $\tau_{\tilde{\chi}}=10$~ns.
The solid and dashed lines show the cuts with and without EMTiming
system usage respectively that give the smallest 95\%~C.L. cross
section limit. In (c) the mass is so large that there is only small
additional acceptance from allowing large $\Delta s_{12}$ events
which is why the sensitivity is not improved in this mass region (see \ref{cap:xsectionsN1-gammagamma}).}
\end{figure*}

A $\gamma\gamma$ + $E_{T}\!\!\!\!\!\!\!/\,\,\,$ analysis is expected
to have the best sensitivity for low neutralino lifetimes. We follow
the analysis in \cite{key-10} (summarized in Table~\ref{cap:gammagammaMETtable}) and allow events
in which both photons have \textcolor{black}{$E_{T}>12$~GeV and
$|\eta|$~<~1}, and study final selection requirements on $E_{T}\!\!\!\!\!\!\!/\,\,\,$ and $\Delta s$. The background for this analysis consists of QCD events with fake
$E_{T}\!\!\!\!\!\!\!/\,\,\,$~\cite{key-31}. We model
the $E_{T}\!\!\!\!\!\!\!/\,\,\,$ from QCD with a resolution of 10~GeV,
i.e. we assume a measurement uncertainty of the transverse energy
of all particles of 10~GeV in each $x$- and $y$-direction, as this
reproduces well the numbers in~\cite{key-10} and allows us to extend
our search region to large values of $E_{T}\!\!\!\!\!\!\!/\,\,\,$. Since all photons from QCD are promptly produced,
we model them with a mean $\Delta s$ of~0~ns, and a resolution
of 1.0~ns.
\textcolor{black}{We found that adding the $\Delta s$ values, $\Delta s_{12}=\Delta s_{1}+\Delta s_{2}$,
and selecting signal events with} \textcolor{black}{\emph{either}}
\textcolor{black}{large} $E_{T}\!\!\!\!\!\!\!/\,\,\,$ \textcolor{black}{}\textcolor{black}{\emph{or}}
\textcolor{black}{large $\Delta s_{12}$, either of which is not SM-like,
maximizes the separation of signal and background as shown in Fig.~\ref{cap:SigBkgdistribution}. The position of
the cuts are optimized for each mass and lifetime to minimize the
95\%~C.L. cross section limit, and we find that both the $\Delta s_{12}$
and} $E_{T}\!\!\!\!\!\!\!/\,\,\,$ \textcolor{black}{cuts are stable at around 7~ns and 50~GeV for non-zero lifetimes. } Without timing information
we found the optimal $E_{T}\!\!\!\!\!\!\!/\,\,\,$ cut to also be around 50~GeV. 

\subsubsection{$\gamma$ + $E_{T}\!\!\!\!\!\!\!/\,\,\,$ + 0 jets}

\begin{figure}
\begin{center}{\includegraphics[%
  width=8cm,
  keepaspectratio]{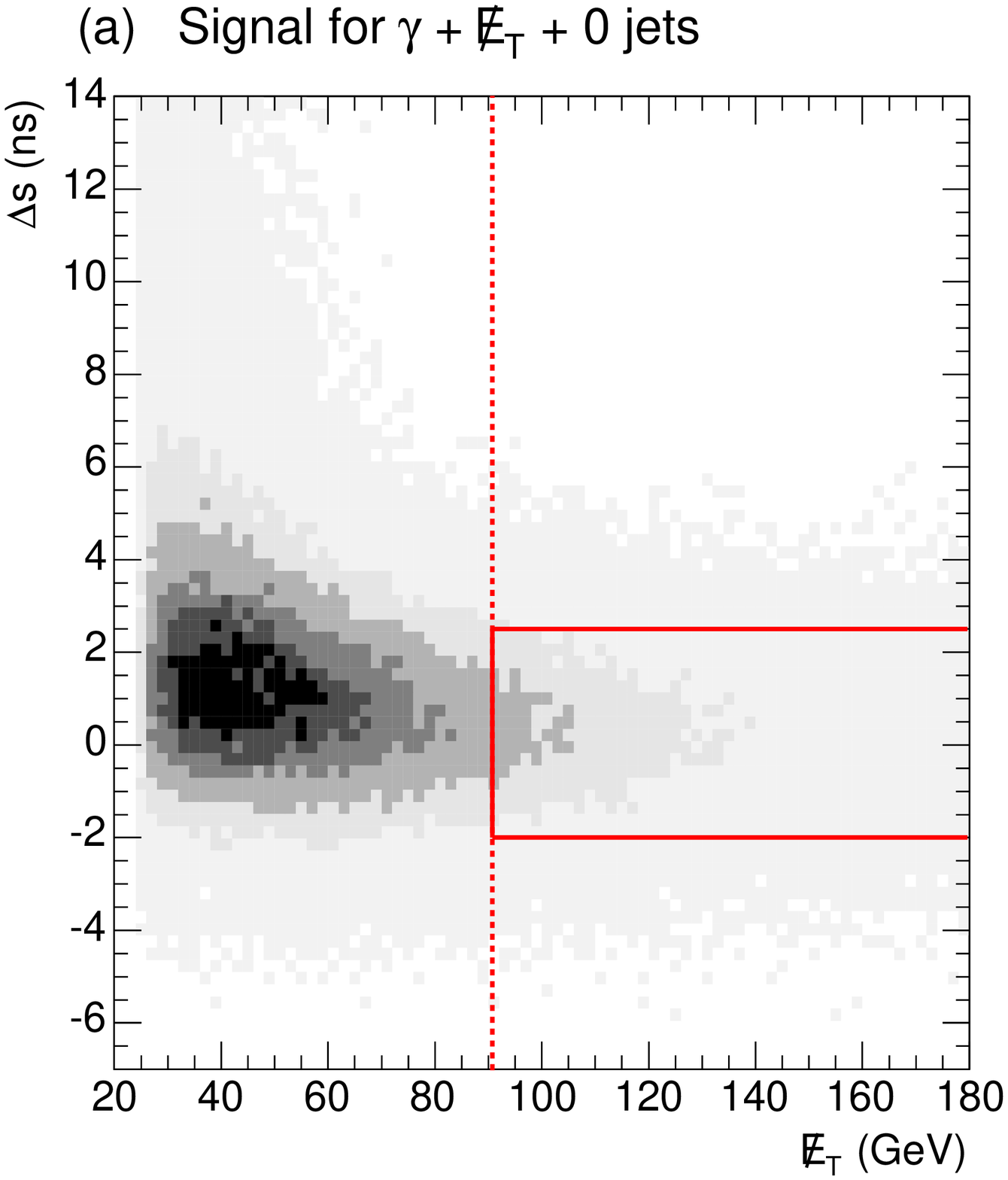}}\end{center}

\begin{center}{\includegraphics[%
  width=8cm,
  keepaspectratio]{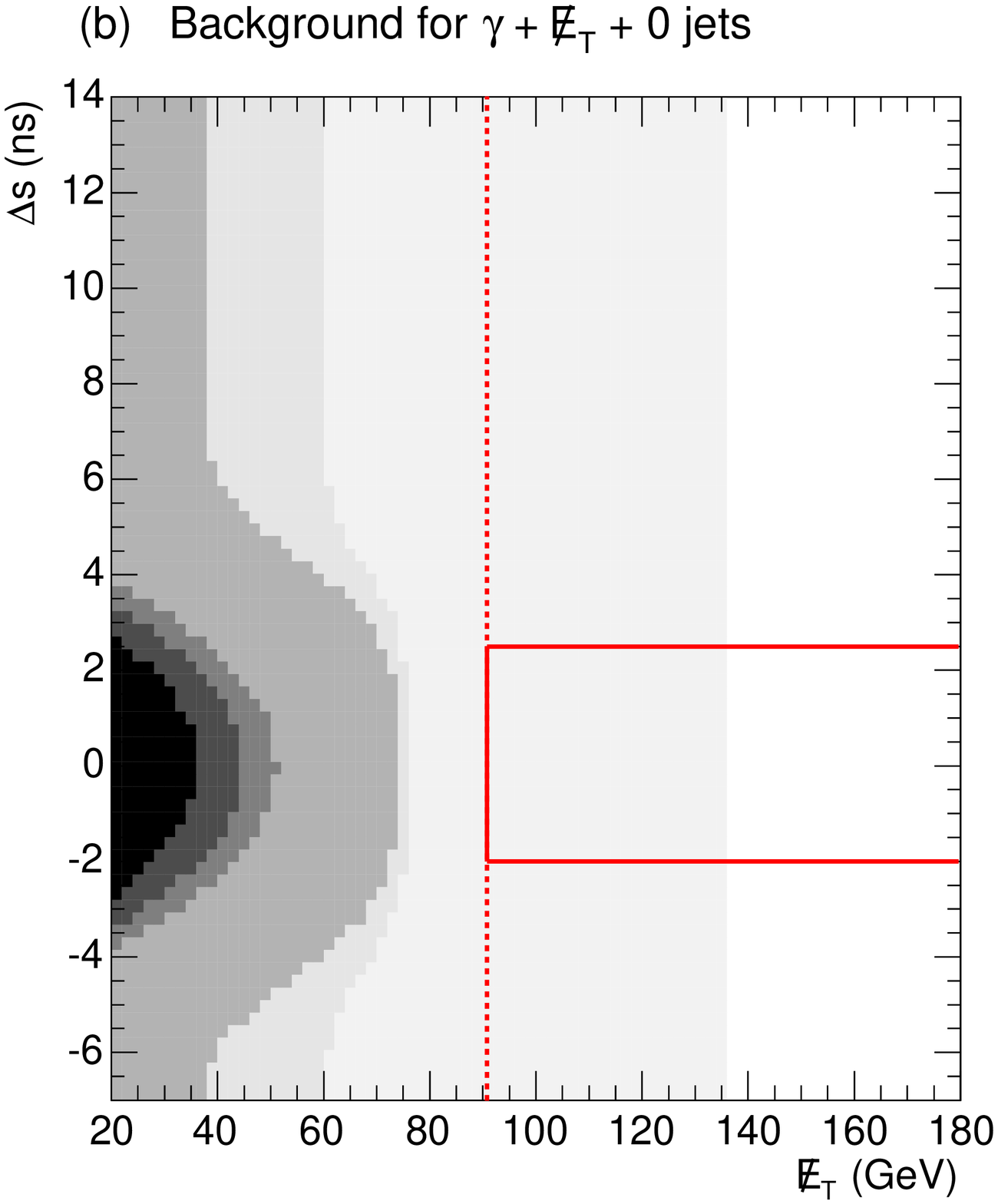}}\end{center}

\caption{\label{cap:SigBkgdistribution_gammaMET0jet}The distribution of $\Delta s$
and $E_{T}\!\!\!\!\!\!\!/\,\,\,$ for signal and background in the $\gamma$~+~$E_{T}\!\!\!\!\!\!\!/\,\,\,$~+~0~jets
analysis. The distributions are (a) from direct neutralino pair production, with $m_{\tilde{\chi}}=70$~\munit\
and $\tau_{\tilde{\chi}}=10$~ns, and (b) from SM background. The
solid and dashed lines show the cuts with and without EMTiming system
usage respectively that give the smallest 95\%~C.L. cross section
limit.}
\end{figure}

From efficiency considerations and since
the signal does not produce jets at the parton level we expect a $\gamma$~+~$E_{T}\!\!\!\!\!\!\!/\,\,\,$~+~0~jets analysis to yield the best sensitivity for longer neutralino lifetimes. We follow the
analysis in~\cite{key-7} (summarized in Table~ \ref{cap:gammaMET0jetstable}) and require the 
highest $E_{T}$ photon to have \textcolor{black}{$E_{T}$~>~25~GeV
and} $|\eta|\le$~1, a minimum $E_{T}\!\!\!\!\!\!\!/\,\,\,$ of 25
GeV, and no jets \textcolor{black}{or additional photons with
$E_{T}$~>~15~GeV}, and study the final selection requirements on  $E_{T}\!\!\!\!\!\!\!/\,\,\,$ and $\Delta s$. 
The background for this analysis is dominantly QCD, $Z\gamma$ and cosmic ray sources.
Since photons from cosmic ray sources hit the
detector with no correlation between the arrival time and the time
of collision, we expect them to be randomly distributed over time
and model this with a flat random distribution in $\Delta s$. As
in the previous section the $\Delta s$ of all other SM sources is
dominated by the timing resolution of 1.0~ns. The $E_{T}\!\!\!\!\!\!\!/\,\,\,$
for the backgrounds are modeled according to the shapes in~\cite{key-7},
and extrapolated to large values using an exponential fit. The expected
background and signal shapes are shown in Fig.~\ref{cap:SigBkgdistribution_gammaMET0jet}.
\textcolor{black}{The final cuts} on $E_{T}\!\!\!\!\!\!\!/\,\,\,$ and
\textcolor{black}{$\Delta s$ are applied} to sort out events with
a large $E_{T}\!\!\!\!\!\!\!/\,\,\,$ \emph{and} whose photon has a
\textcolor{black}{$\Delta s$ within a lower bound and an upper bound:
$\Delta s_{1}\le\Delta s\le\Delta s_{2}$, to reject photons from
SM background as well as from cosmic ray sources. We find the optimized cuts at around $\Delta s_{1}=-2.0$~ns,
$\Delta s_{2}=2.0$~ns and} $E_{T}\!\!\!\!\!\!\!/\,\,\,=80$~GeV.
\textcolor{black}{However, $\Delta s_{2}$ could vary up to 3~ns
for high lifetimes,} $E_{T}\!\!\!\!\!\!\!/\,\,\,$ \textcolor{black}{up
to} 120~GeV for higher \textcolor{black}{masses. Without timing information
we found the} optimal $E_{T}\!\!\!\!\!\!\!/\,\,\,$ cut mostly
at around 100~GeV. 

\begin{table}

\caption{\label{cap:gammagammaMETtable}The background and baseline selection
criteria used for the $\gamma\gamma$~+~$E_{T}\!\!\!\!\!\!\!/\,\,\,$
analysis following Ref.~\cite{key-10,key-31}.}

\begin{center}\textcolor{black}{}\begin{tabular}{p{8.6cm}}
\hline 
\hline

Baseline selection requirements:\tabularnewline
\hline 
\textcolor{black}{$E_{T}^{\gamma_{1}} > 12$~GeV, $E_{T}^{\gamma_{2}} > 12$~GeV}\tabularnewline
$|\eta^{\gamma_{1}}|<1$, $|\eta^{\gamma_{2}}|<1$\tabularnewline
\hline
\hline 
\textcolor{black}{Backgrounds: }\tabularnewline
\hline 
\textcolor{black}{2,577 events / $100\ \mathrm{pb}^{-1}$ from QCD}\tabularnewline
\textcolor{black}{$\Delta s_{12}=\Delta s^{\gamma_{1}}+\Delta s^{\gamma_{2}}$,
$\overline{\Delta s_{12}}=0$~ns with resolution $\sigma_{\Delta s_{12}}=1.41$~ns}\tabularnewline
$E_{T}\!\!\!\!\!\!\!/\,\,\,$\textcolor{black}{: Rayleigh distribution
(Square-root of the sum of 2 Gaussians squared) with $\sigma$~=~10~GeV}\tabularnewline
\hline
\hline 
\textcolor{black}{Optimization: }\tabularnewline
\hline
\textcolor{black}{Accept events where the event has a }
$E_{T}\!\!\!\!\!\!\!/\,\,\,$ \textcolor{black}{greater
than the optimized $E_{T}\!\!\!\!\!\!\!/\,\,\,$ cut}
\textcolor{black}{\emph{or}} \textcolor{black}{whose photon has a $\Delta s_{12}$
greater than the optimized $\Delta s_{12}$ cut. }\tabularnewline
\hline
\hline
\end{tabular}\end{center}
\end{table}
\begin{table}

\caption{\label{cap:gammaMET0jetstable}The background and baseline selection
criteria for the $\gamma$~+ $E_{T}\!\!\!\!\!\!\!/\,\,\,$~+~0~jets
analysis following Ref.~\cite{key-7}.}

\begin{center}\textcolor{black}{}\begin{tabular}{p{8.6cm}}
\hline 
\hline

Baseline selection requirements:\tabularnewline
\hline 
\textcolor{black}{$E_{T}^{\gamma}$ > 25 GeV}\tabularnewline
$|\eta^{\gamma}|$ < 1.0\tabularnewline
$E_{T}\!\!\!\!\!\!\!/\,\,\,$ \textcolor{black}{> 25 GeV}\tabularnewline
\textcolor{black}{No jets or additional photons with $E_{T}^{jet}>15$ GeV}\tabularnewline
\hline
\hline 
\textcolor{black}{Backgrounds: }\tabularnewline
\hline 
\textcolor{black}{12.6 events / $100\ \mathrm{pb}^{-1}$ from $Z\gamma\rightarrow\nu\bar{\nu}\gamma$,
W$\gamma$, \mbox{$W\rightarrow e\nu$,} QCD and cosmics} \tabularnewline
\textcolor{black}{Non-cosmics: $\Delta s=0$, $\overline{\Delta s_{12}}=0$~ns
with resolution $\sigma_{\Delta s}=1.0$~ns}\tabularnewline
\textcolor{black}{Cosmics: 57.2\% of total background, flat distribution
in $\Delta s$}\tabularnewline
$E_{T}\!\!\!\!\!\!\!/\,\,\,$ \textcolor{black}{distribution according
to \cite{key-7}, and extrapolated using an exponential}\tabularnewline
\hline
\hline 
Optimization:\tabularnewline
\hline
Accept events where the event has a $E_{T}\!\!\!\!\!\!\!/\,\,\,$ greater
than the optimized cut $E_{T}\!\!\!\!\!\!\!/\,\,\,$ \emph{and} whose photon has a \textcolor{black}{$\Delta s$
within a range of optimized cuts $\Delta s_{1}\le\Delta s\le\Delta s_{2}$.}\tabularnewline
\hline
\hline
\end{tabular}\end{center}
\end{table}

\subsubsection{Results}

\begin{figure}
\begin{center}{\includegraphics[%
  width=8.6cm,
  keepaspectratio]{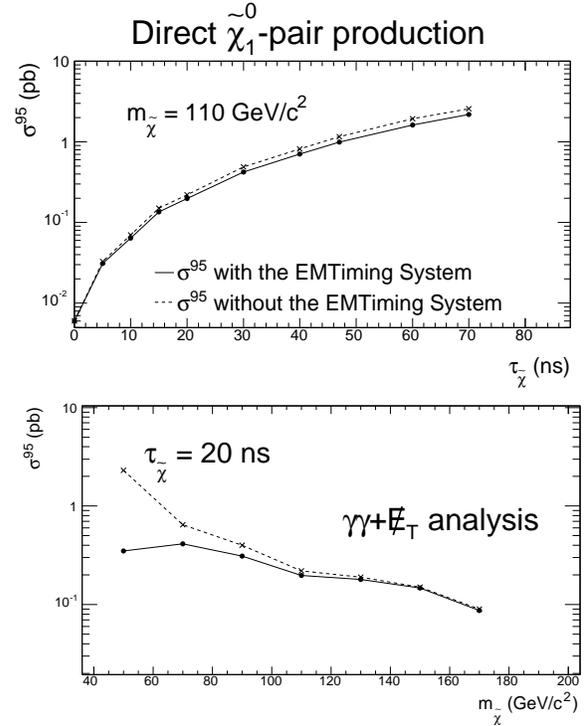}}\end{center}

\caption{\textcolor{black}{\label{cap:xsectionsN1-gammagamma}The expected 95\%~C.L.
cross sections limits on neutralino pair production
as a function of $\tau_{\tilde{\chi}}$ for $m_{\tilde{\chi}}$~=~70~}\munit\
\textcolor{black}{(top) and as a function of $m_{\tilde{\chi}}$ for
$\tau_{\tilde{\chi}}$~=~20~ns (bottom) in the} $\gamma\gamma$~+~$E_{T}\!\!\!\!\!\!\!/\,\,\,$
analysis \textcolor{black}{for $2\ \mathrm{fb}^{-1}$ luminosity both
with and without a timing system for comparison. As expected at $\tau_{\tilde{\chi}}=0$~ns
the cross sections merge as the timing system has no effect; for higher
$\tau_{\tilde{\chi}}$ the sensitivity goes down as more photons leave
the detector, but the difference of the limits increases as $\Delta s$
gets larger for the signal and timing becomes more helpful. The limits
get better as the mass goes up since more of the events pass the kinematic
requirements, however the timing system only provides real additional
sensitivity at the lowest masses where the neutralino momentum distribution
is softer.}}
\end{figure}

\begin{figure}
\begin{center}\includegraphics[%
  width=8.6cm,
  keepaspectratio]{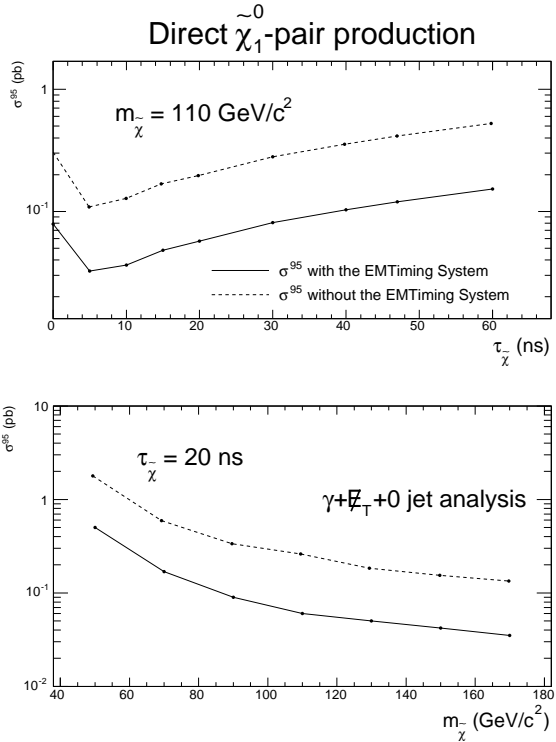}\end{center}

\caption{\textcolor{black}{\label{cap:xsectionsN1-gamma}The expected 95\% C.L. cross
section limits on neutralino pair production as a function of $\tau_{\tilde{\chi}}$ for $m_{\tilde{\chi}}$~=~70~}\munit\
\textcolor{black}{(top) and as a function of $m_{\tilde{\chi}}$ for $\tau_{\tilde{\chi}}$~=~20~ns
(bottom) in} the \textcolor{black}{$\gamma$~+}~$E_{T}\!\!\!\!\!\!\!/\,\,\,$\textcolor{black}{~+~0
jets analysis for $2\ \mathrm{fb}^{-1}$ both with and without a timing
system. As in Fig.~\ref{cap:xsectionsN1-gammagamma}, for higher
$\tau_{\tilde{\chi}}$ the sensitivity goes down as more photons leave
the detector, but the difference of the limits increases as $\Delta s$
gets larger for the signal and timing becomes more helpful. The curves
do not merge at 0~ns lifetime since cosmic ray backgrounds always contribute
at high $\Delta s$ and the timing system always has some effect on
the cross section limit. The rise from 10~ns to 0~ns originates
in an increasing probability towards zero lifetime for two photons
to remain in the detector, yielding lower efficiency for a single photon
analysis. The limits get better as the mass goes up since more of
the events pass the kinematic requirements.}}
\end{figure}

\begin{figure}
\begin{center}\includegraphics[%
  width=8.6cm]{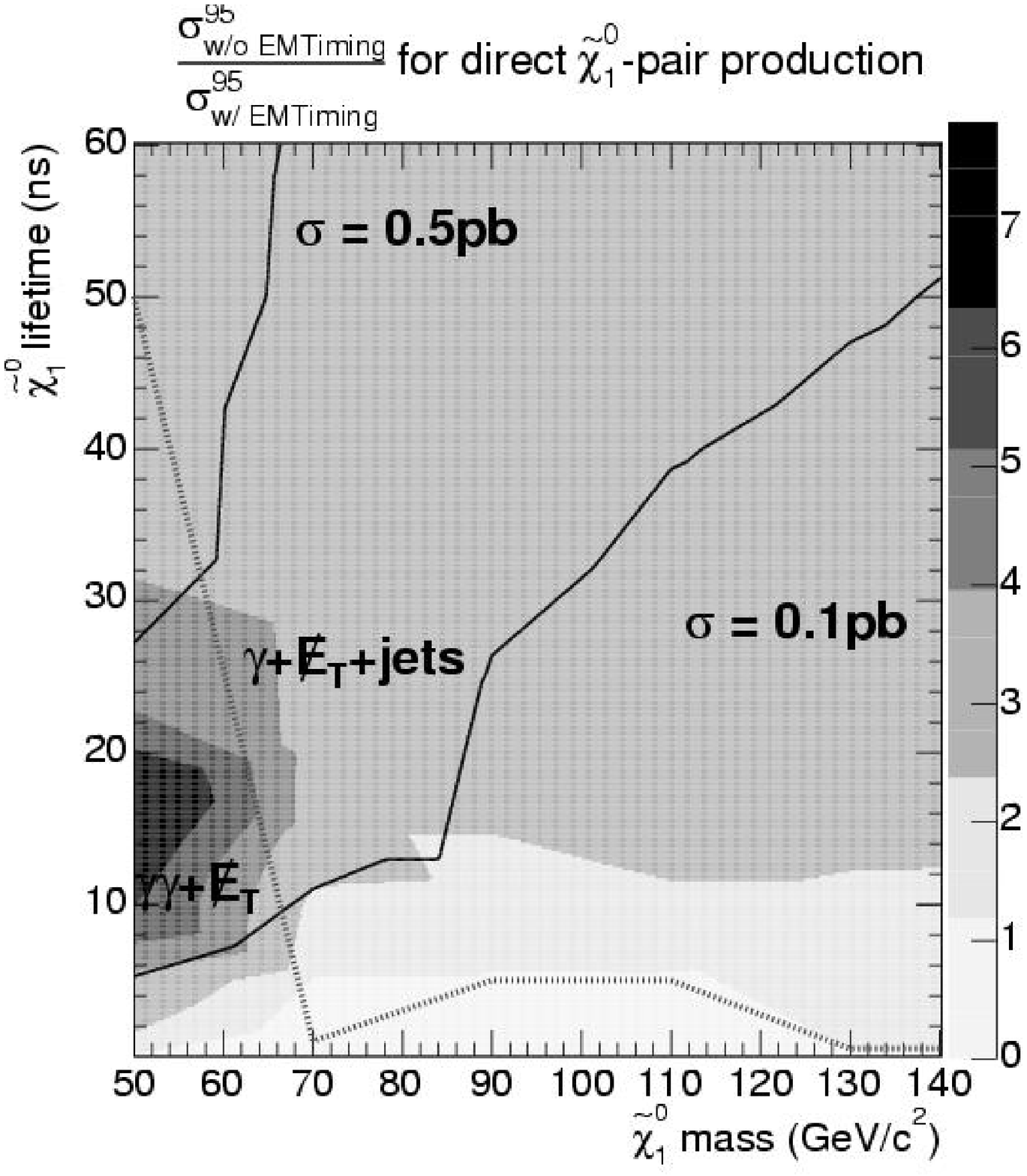}\end{center}

\caption{\label{cap:emtiming-difference-N1N1}This plot combines the $\gamma\gamma$~+~$E_{T}\!\!\!\!\!\!\!/\,\,\,$
and $\gamma$~+~$E_{T}\!\!\!\!\!\!\!/\,\,\,$~+~0~jets
analysis results for neutralino pair production for
\textcolor{black}{$2\ \mathrm{fb}^{-1}$} of data and is \textcolor{black}{a
2-dimensional visualization of Figs.~\ref{cap:xsectionsN1-gammagamma}
and \ref{cap:xsectionsN1-gamma}.} The contours of constant cross
section limit are shown as the solid lines, and the separation line
between the regions where the two different analyses provide the best
sensitivity is given by the dotted line; \textcolor{black}{the $\gamma$~+}~$E_{T}\!\!\!\!\!\!\!/$
\textcolor{black}{+~0~jets analysis shows better cross section
limits than} a $\gamma\gamma$~+~$E_{T}\!\!\!\!\!\!\!/\,\,\,$ analysis
\textcolor{black}{in the mass and lifetime range above the dashed
line.} The shaded regions delineate the contours of the ratio of \textcolor{black}{the
95\%~C.L. cross section limits between with and without EMTiming
information. The ratio is greatest for a low neutralino mass and a
lifetime of 10-20~ns, and lowest for a high mass and low lifetime. }}
\end{figure}

Figures~\ref{cap:xsectionsN1-gammagamma} and \ref{cap:xsectionsN1-gamma}
show the expected 95\% C.L. cross section upper limits vs. \textcolor{black}{$\tau_{\tilde{\chi}}$
for $m_{\tilde{\chi}}$~=~70~}\munit\ \textcolor{black}{and vs.
$m_{\tilde{\chi}}$ for $\tau_{\tilde{\chi}}$~=~20~ns for both
analyses for a luminosity of $2\ \mathrm{fb}^{-1}$.} Figure~\ref{cap:emtiming-difference-N1N1}
shows the ratio of the \textcolor{black}{lower of the two 95\%~C.L.
cross section limits with EMTiming system usage and without in two
dimensions.} In these plots we see four trends: 1) As a function of
lifetime the \textcolor{black}{cross section limits rise} since the
probability that the neutralinos decay in the detector goes down,
\textcolor{black}{2) at high lifetimes the timing handle is better
able to separate the signal from the backgrounds and produces better
limits relative to kinematics alone, 3) as a function of mass the
cross section decreases as more and more events pass the kinematic
threshold, and 4) at low masses the timing handle is better able to
separate the signal from the backgrounds because the momentum distribution
happens be lower on average (see Fig.~\ref{cap:SigBkgdistribution}}c
which shows an example where the kinematics are such that there is
only small additional acceptance from allowing large $\Delta s_{12}$
events). As expected the $\gamma\gamma$~+~$E_{T}\!\!\!\!\!\!\!/\,\,\,$ analysis
\textcolor{black}{yields lower cross sections when the mass or the
lifetime is low (see Fig.~\ref{cap:emtiming-difference-N1N1}). The ratio is greatest in this region and occurs at
a mass around 50~}\munit\ \textcolor{black}{and a lifetime of 10-20~ns.}
The \textcolor{black}{$\gamma$~+}~$E_{T}\!\!\!\!\!\!\!/\,\,\,$\textcolor{black}{~+~0~jets
analysis yields lower cross section limits for the rest of the considered
lifetime range and masses, but it is important to note that the course
of the separation line of the analyses depends on the production momentum
distribution of the neutralino. Unfortunately this analysis cannot
be applied to searches for long-lived NLSP neutralinos in a true GMSB
model with the preferred production processes as there the neutralinos
are produced as part of cascades from gaugino pairs which contain
jets. Therefore, we do a separate analysis for a full GMSB production
in the next section.}



\section{Sensitivity to GMSB models}

We next consider the sensitivity to full GMSB production 
where we allow all processes to contribute to the final
state according to their predicted cross sections. We use the same
simulation tools as in Section \ref{sub:Analysis-Methods-and}, with
the GMSB parameters chosen according to the Snowmass Slope guidelines~\cite{key-23} in the range where the neutralino
is the NLSP. Again we consider a single photon and a diphoton analysis.
The $\gamma\gamma$~+~$E_{T}\!\!\!\!\!\!\!/\,\,\,$ analysis methodology
is identical to the case of neutralino pair production, but the single photon
analysis must be modified to allow jets as here the neutralinos are
part of cascades from gauginos which produce additional particles
which, in general, could be identified jets. We thus use a $\gamma$~+~$E_{T}\!\!\!\!\!\!\!/\,\,\,$~+~jets
analysis.

\subsection{$\gamma$ + $E_{T}\!\!\!\!\!\!\!/\,\,\,$ + jets }

\begin{figure}
\begin{center}{\includegraphics[%
  height=8cm,
  keepaspectratio]{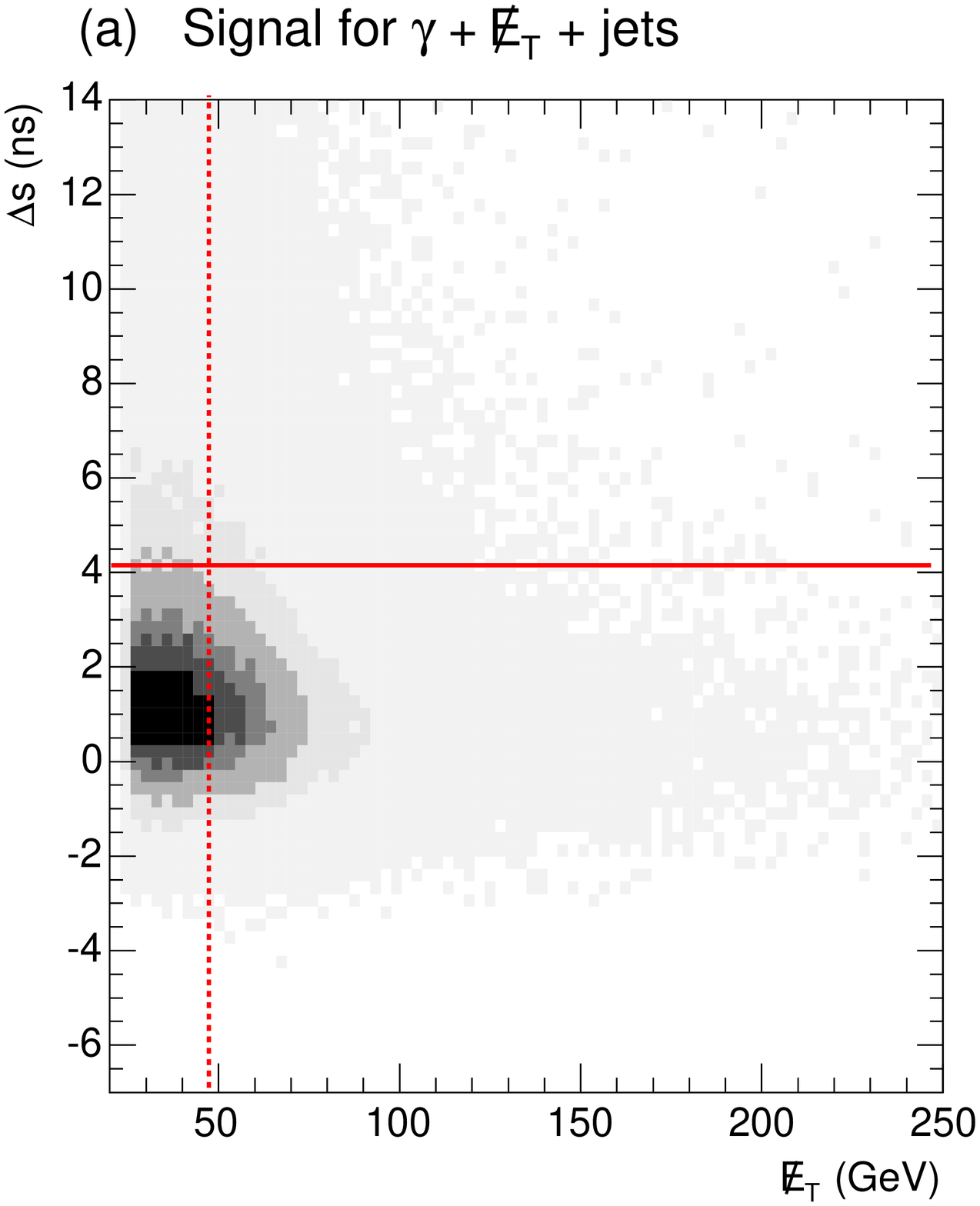}}\end{center}

\begin{center}{\includegraphics[%
  height=8cm,
  keepaspectratio]{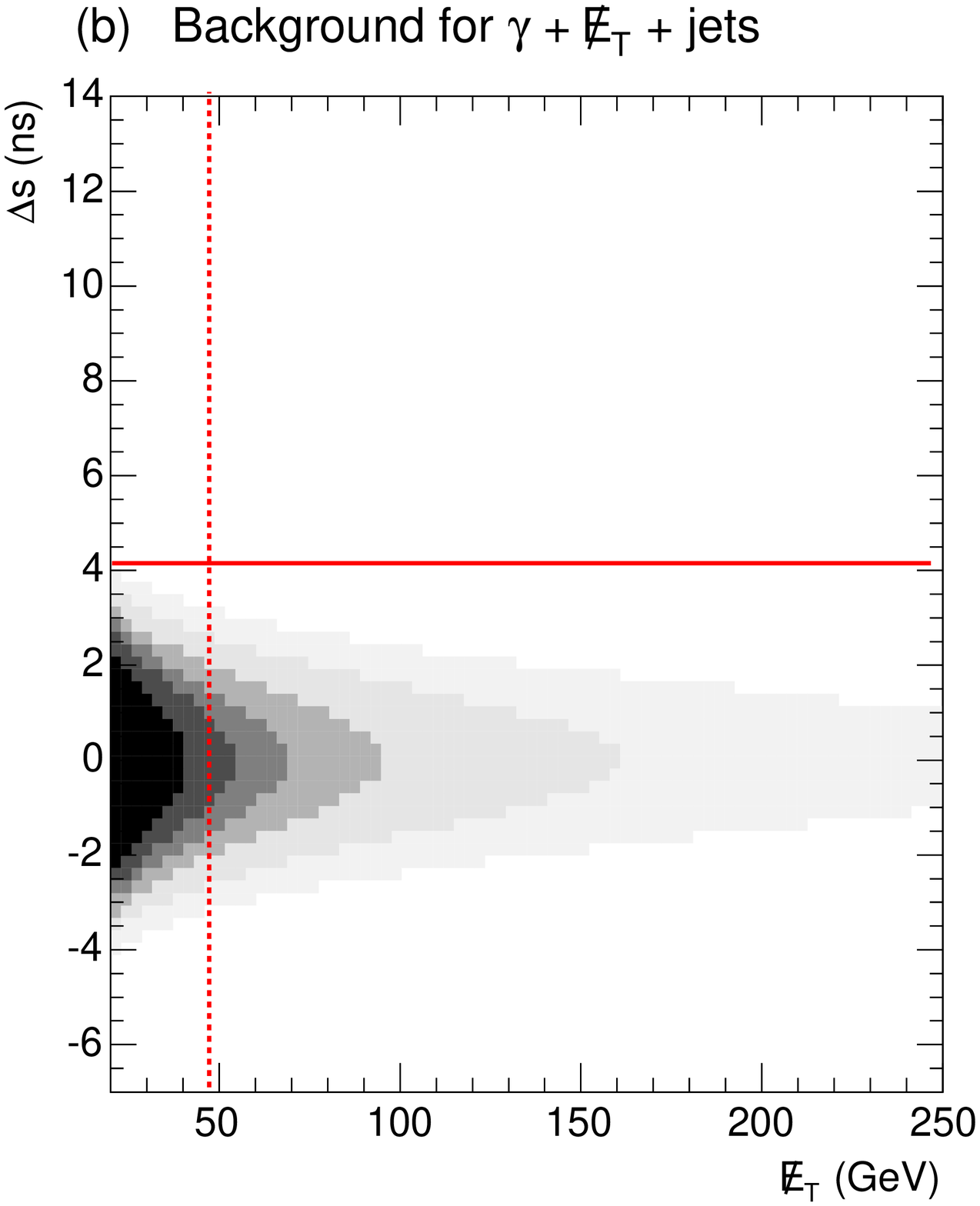}}\end{center}

\caption{\label{cap:SigBkgdistribution_gammaMETjets}The distribution of $\Delta s$
and $E_{T}\!\!\!\!\!\!\!/\,\,\,$ for signal and background in the $\gamma$~+~$E_{T}\!\!\!\!\!\!\!/\,\,\,$~+~jets
analysis. The distributions are (a) from full GMSB
production, with $m_{\tilde{\chi}}=70$~\munit\
and $\tau_{\tilde{\chi}}=10$~ns, and (b) from SM background. The
solid and dashed lines show the cuts with and without EMTiming system
usage respectively that give the smallest 95\%~C.L. cross section
limit.}
\end{figure}

A $\gamma$~+~$E_{T}\!\!\!\!\!\!\!/\,\,\,$~+~jets analysis should
be most sensitive to neutralinos with long lifetime which are produced
in association with other particles in the final state such as from
gaugino pair production. We follow the analysis in~\cite{key-8} (summarized in Table~\ref{cap:gammaMETjetstable})
and require events with the primary (highest $E_{T}$) photon
to have \textcolor{black}{$E_{T}>$~25~GeV and} $|\eta|$~=~{[}0,
1.1{]} or {[}1.5, 2.0{]}, $E_{T}\!\!\!\!\!\!\!/\,\,\,>25$~GeV, $\geq$~2~jets
of \textcolor{black}{$E_{T}$~>~20~GeV} and \textcolor{black}{$|\eta|$~<~2.0}, and study the final selection requirements on  $E_{T}\!\!\!\!\!\!\!/\,\,\,$ and $\Delta s$. The backgrounds are dominated by QCD and W+jets~\cite{key-34}. 
\textcolor{black}{The expected} $E_{T}\!\!\!\!\!\!\!/\,\,\,$ \textcolor{black}{of
the background is modeled according to} Ref.~\cite{key-8}\textcolor{black}{,
and since the backgrounds are from SM we take a mean $\Delta s=0$~ns}
with a resolution of 1.0~ns. \textcolor{black}{The signal and background
shapes are shown in Fig.~\ref{cap:SigBkgdistribution_gammaMETjets}.}
\textcolor{black}{We find that the optimal
final selection requirements accept} events in which the event has
a large $E_{T}\!\!\!\!\!\!\!/\,\,\,$ \emph{or} a large \textcolor{black}{$\Delta s$.
Again, both cuts are optimized to minimize the 95\%~C.L. cross section
limit for each mass and lifetime case. For without-timing usage we
find the} $E_{T}\!\!\!\!\!\!\!/\,\,\,$ cut to \textcolor{black}{be
around 50~GeV for masses around 70~}\munit\textcolor{black}{, varying
up to 110~GeV for masses around 150~}\munit\textcolor{black}{.
For with-timing usage we find only a $\Delta s$ cut around 4~ns
which is stable for all masses and lifetimes, and no} $E_{T}\!\!\!\!\!\!\!/\,\,\,$
\textcolor{black}{cut other than the baseline} $E_{T}\!\!\!\!\!\!\!/\,\,\,$~>~25~GeV
\textcolor{black}{(except for} $\tau_{\tilde{\chi}}=0$~ns \textcolor{black}{where
the diphoton case has the best sensitivity).
While it is outside of our ability to predict, one might find further
optimization is possible by further lowering the baseline selection
requirements. }

\begin{table}

\caption{\label{cap:gammaMETjetstable}The background and baseline selection criteria
used for the $\gamma$ + $E_{T}\!\!\!\!\!\!\!/\,\,\,$ + jets analysis following Refs.~\cite{key-8,key-34}.}

\begin{center}\textcolor{black}{}\begin{tabular}{p{8.6cm}}
\hline 
\hline 

Baseline selection requirements:\tabularnewline
\hline
\textcolor{black}{$E_{T}^{\gamma}>20$ GeV}\tabularnewline
\textcolor{black}{$|\eta^{\gamma}|<1.1$ and $1.5<|\eta^{\gamma}|<2.0$}\tabularnewline
$E_{T}\!\!\!\!\!\!\!/\,\,\,$ \textcolor{black}{> 25 GeV}\tabularnewline
\textcolor{black}{$\geq$2 jets with $E_{T}^{jet}$ > 20 GeV and $|\eta^{jet}|$~<~2.0}\tabularnewline
\hline
\hline 
\textcolor{black}{Backgrounds:}\tabularnewline
\hline 
 \textcolor{black}{320~events / $100\ \mathrm{pb}^{-1}$ from QCD
and W+jets}\tabularnewline
\textcolor{black}{$\overline{\Delta s_{12}}=0$~ns with resolution
$\sigma_{\Delta s}=1.0$~ns}\tabularnewline
$E_{T}\!\!\!\!\!\!\!/\,\,\,$ \textcolor{black}{distribution from \cite{key-8},
extrapolated to large} $E_{T}\!\!\!\!\!\!\!/\,\,\,$\tabularnewline
\hline
\hline 
Optimization:\tabularnewline
\hline
\textcolor{black}{Accept events where the event has}
\textcolor{black}{a} $E_{T}\!\!\!\!\!\!\!/\,\,\,$ \textcolor{black}{greater
than the optimized $E_{T}\!\!\!\!\!\!\!/\,\,\,$ cut} \textcolor{black}{\emph{or}} \textcolor{black}{whose photon has a $\Delta s$
greater than the optimized $\Delta s$ cut.}\tabularnewline
\hline
\hline
\end{tabular}\end{center}
\end{table}

\subsection{Results }

\begin{figure}
\begin{center}\includegraphics[%
  width=8.6cm,
  keepaspectratio]{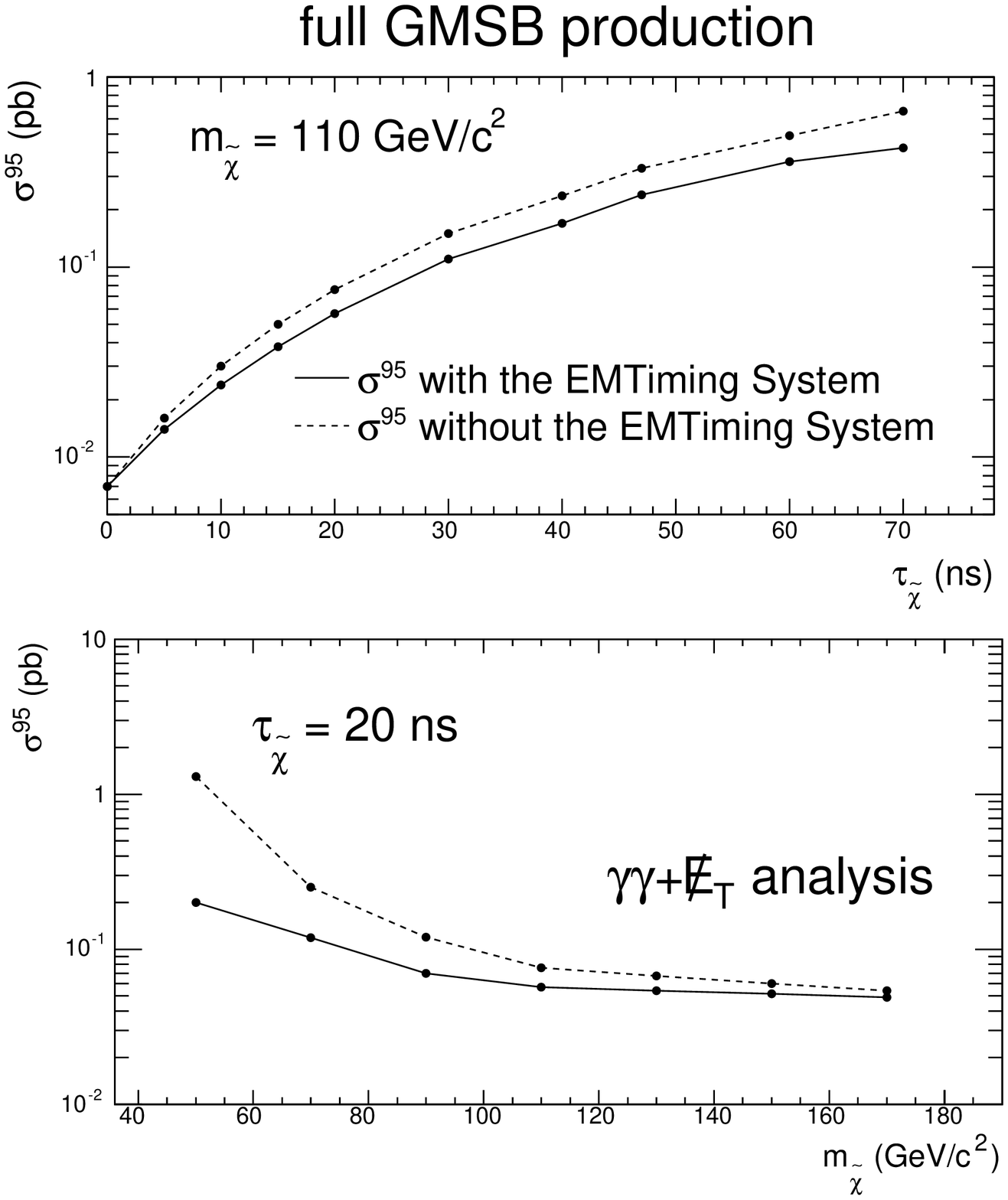}\end{center}

\caption{\textcolor{black}{\label{cap:xsectionsgmsb-gammagamma}The expected 95\%~C.L.
cross section limits as a function of} $\tau_{\tilde{\chi}}$ \textcolor{black}{for
$m_{\tilde{\chi}}$~=~70} \munit\ \textcolor{black}{(top) and as a function of
$m_{\tilde{\chi}}$ for} $\tau_{\tilde{\chi}}$\textcolor{black}{~=~20~ns
(bottom) at $2\ \mathrm{fb}^{-1}$ for with and without EMTiming for
full GMSB production in a $\gamma\gamma$~+}~$E_{T}\!\!\!\!\!\!\!/\,\,\,$
analysis. The results are similar to those in Fig.~\ref{cap:xsectionsN1-gammagamma}.}
\end{figure}

\begin{figure}
\begin{center}\includegraphics[%
  width=8.6cm,
  keepaspectratio]{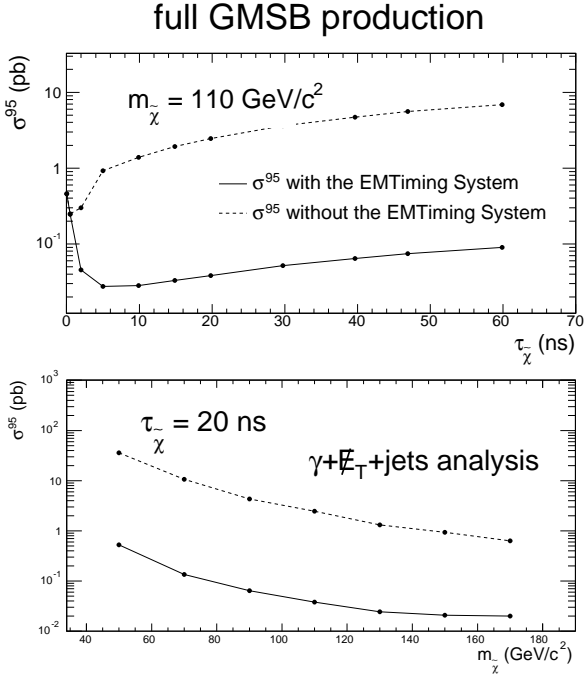}\end{center}

\caption{\textcolor{black}{\label{cap:xsections-gmsb-gammaMET}The expected 95\%~C.L.
cross section limits as a function of} $\tau_{\tilde{\chi}}$ \textcolor{black}{for
$m_{\tilde{\chi}}$~=~70~}\munit\ \textcolor{black}{(top) and as a function of
$m_{\tilde{\chi}}$ for} $\tau_{\tilde{\chi}}$\textcolor{black}{~=~20~ns
(bottom) at $2\ \mathrm{fb}^{-1}$ for with and without EMTiming System
for full GMSB production in a $\gamma$~+}~$E_{T}\!\!\!\!\!\!\!/\,\,\,$\textcolor{black}{~+~jets
analysis. For all but the lowest lifetimes the timing information
significantly improves the cross section limits. Note that here the
cross sections merge at zero lifetime since we have neglected cosmics
in this analysis following} \cite{key-8, key-34}\textcolor{black}{.}}
\end{figure}

\begin{figure}
\begin{center}\includegraphics[%
  width=8.6cm]{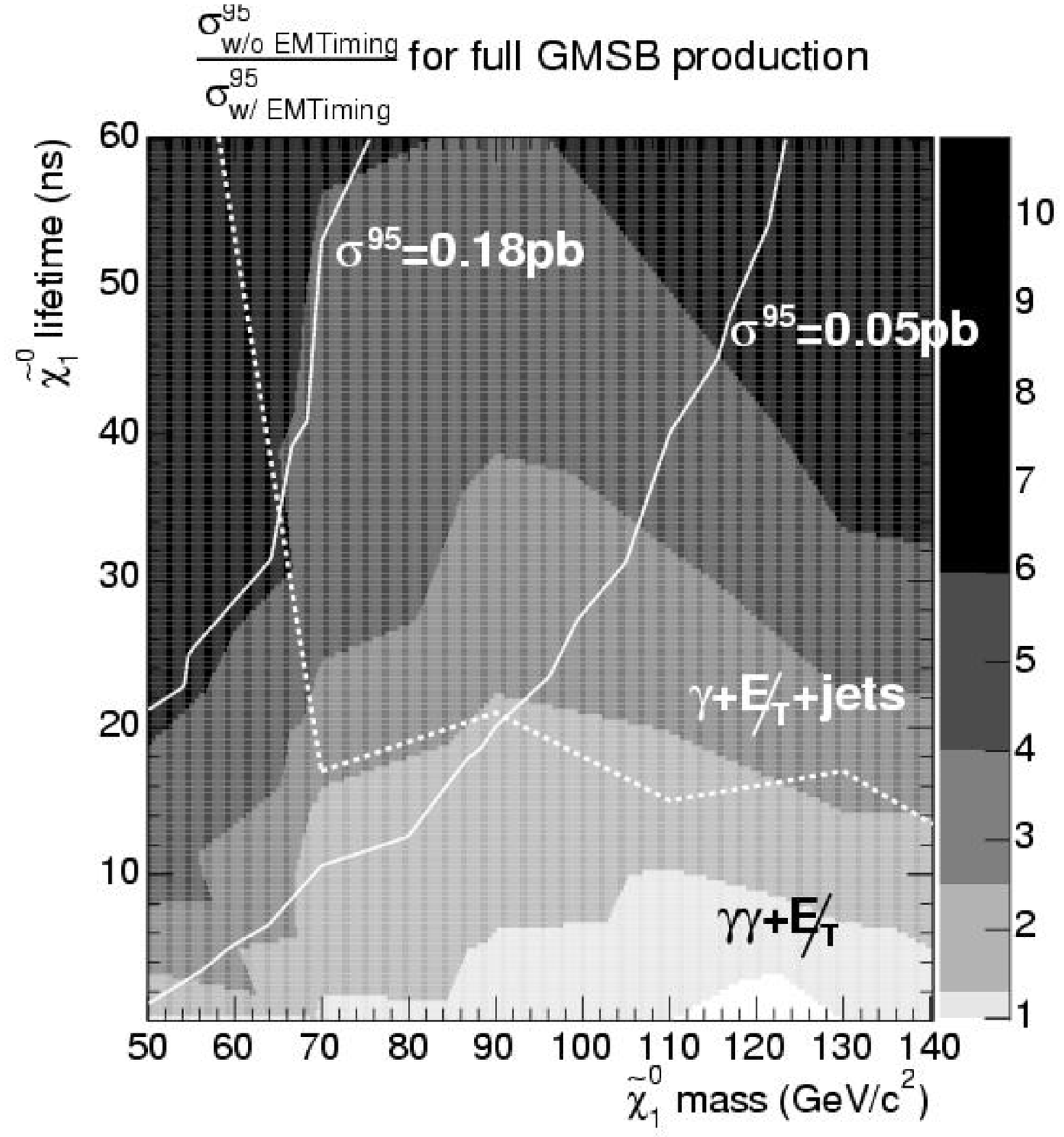}\end{center}

\caption{\label{cap:emtiming-difference-gmsb}This plot combines the $\gamma\gamma$~+~$E_{T}\!\!\!\!\!\!\!/\,\,\,$
and \textcolor{black}{$\gamma$~+}~$E_{T}\!\!\!\!\!\!\!/\,\,\,$\textcolor{black}{~+~jets
analysis results for a full GMSB model simulation} for \textcolor{black}{$2\ \mathrm{fb}^{-1}$}
of data and is \textcolor{black}{a 2-dimensional visualization of
Figs. \ref{cap:xsectionsgmsb-gammagamma} and \ref{cap:xsections-gmsb-gammaMET}.}
The contours of constant cross section limit are shown as the solid
lines, and the separation line between the regions where the two different
analyses provide the best sensitivity is given by the dotted line;
\textcolor{black}{the $\gamma$~+}~$E_{T}\!\!\!\!\!\!\!/\,\,\,$ \textcolor{black}{+~jets
analysis shows better cross section limits than} a $\gamma\gamma$~+~$E_{T}\!\!\!\!\!\!\!/\,\,\,$
analysis \textcolor{black}{in the mass and lifetime range above the
dashed line.} The shaded regions delineate the contours of constant
ratio of \textcolor{black}{the 95\% C.L.- cross-section limits between
with and without EMTiming information. The EMTiming system has its
most effective region at high lifetime and is least effective for
high masses, where the kinematics give the best separation. }}
\end{figure}

Figures~\ref{cap:xsectionsgmsb-gammagamma} and \ref{cap:xsections-gmsb-gammaMET}
show the expected 95\%~C.L. cross section upper limits vs. \textcolor{black}{$\tau_{\tilde{\chi}}$
for $m_{\tilde{\chi}}$~=~70~}\munit\ \textcolor{black}{and vs.
$m_{\tilde{\chi}}$ for $\tau_{\tilde{\chi}}$~=~20~ns for both
analyses for a luminosity of $2\ \mathrm{fb}^{-1}$.} Figure~\textcolor{black}{\ref{cap:emtiming-difference-gmsb}}
shows the ratio of the \textcolor{black}{lowest 95\%~C.L. cross section
limits with EMTiming system usage and without in two dimensions. We
see the same general trends as in neutralino pair production as the
signal shapes are similar in both analyses. }

\begin{figure}
\begin{center}{\includegraphics[%
  width=8.6cm,
  keepaspectratio]{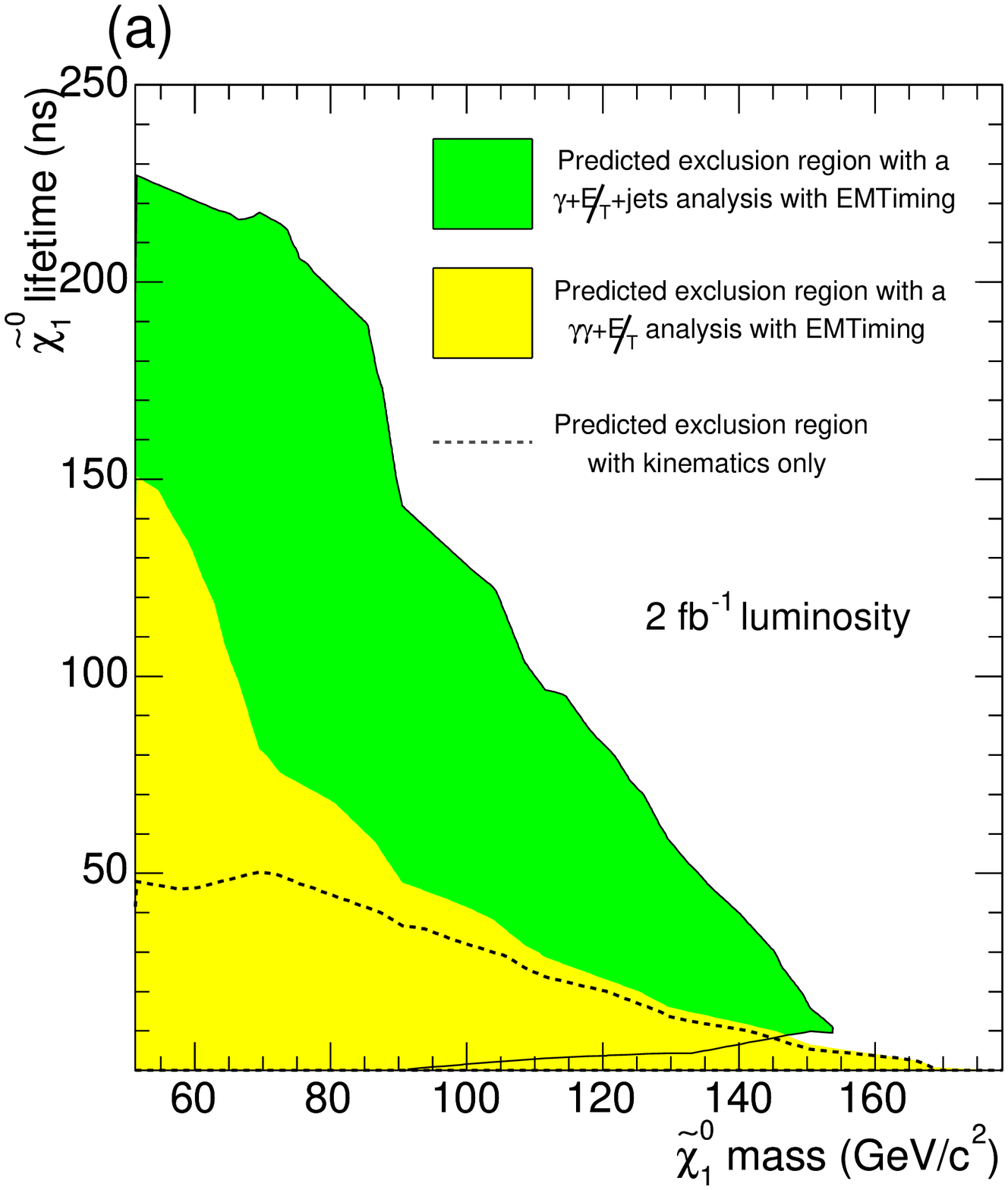}}\end{center}

\begin{center}{\includegraphics[%
  width=8.6cm,
  keepaspectratio]{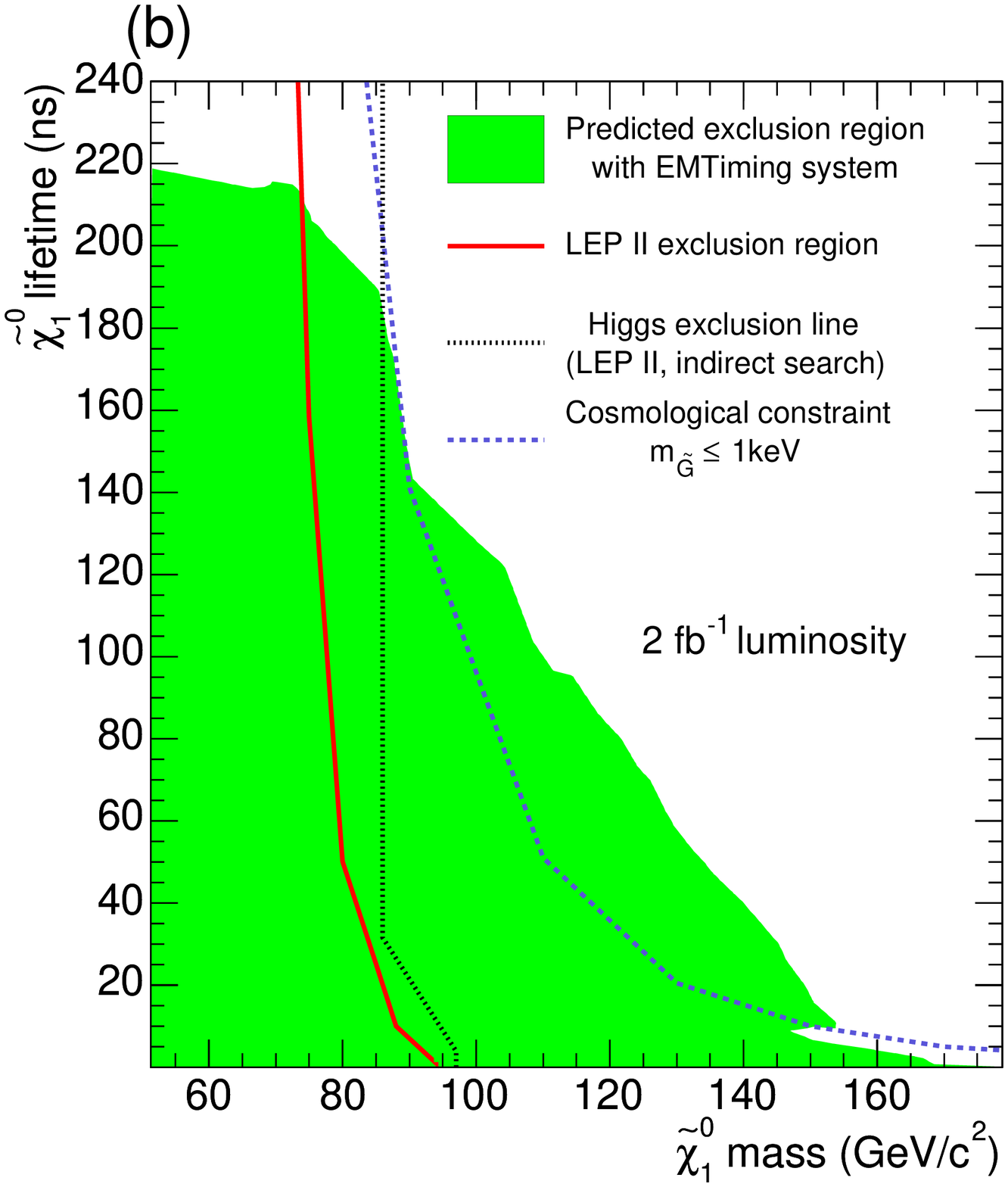}}\end{center}

\caption{\label{cap:xsecmin-gmsb-2000lumi}\textcolor{black}{The expected 95\%~C.L.} exclusion regions as a function of neutralino lifetime
and mass for full GMSB production at \textcolor{black}{$2\ \mathrm{fb}^{-1}$}
luminosity for the \textcolor{black}{the $\gamma\gamma$~+}~$E_{T}\!\!\!\!\!\!\!/\,\,\,$
\textcolor{black}{and a $\gamma$~+}~$E_{T}\!\!\!\!\!\!\!/\,\,\,$\textcolor{black}{~+~jets
analysis} separately. In plot (a) the region below the dashed line is the exclusion
region from kinematics alone, i.e., where no timing information is
used. Plot (b) shows the full exclusion region from the overlap of the two
analyses and compares the results to the direct and indirect search limits from LEP~II~\cite{key-3} 
and the theoretically favored region from
cosmological constraints ($m_{\tilde{G}}<1$~$\mathrm{keV}/c^2$)~\cite{key-27}. The Tevatron in run~II should be able to significantly extend the LEP~II limits and provide sensitivity in the favored region for all masses below about 150~\munit.}
\end{figure}

\textcolor{black}{A comparison of the cross section limits with the
production cross sections in the GMSB model at a luminosity of $2\ \mathrm{fb}^{-1}$
gives the mass vs. lifetime exclusion regions shown in Fig.~\ref{cap:xsecmin-gmsb-2000lumi}.
As expected, timing has the biggest effect at low masses and high lifetimes.
We have also indicated the exclusion regions from LEP~II from both direct
and indirect searches~\cite{key-3}, and the line} $m_{\tilde{G}}=1$~$\mathrm{keV}/c^2$,
below which is the theoretically preferred region from cosmological
constraints~\cite{key-27}. LEP effectively excludes all \textcolor{black}{neutralino
masses under 80~}\munit\ \textcolor{black}{up to high lifetimes,
with a small extension to 100~}\munit\ \textcolor{black}{for lifetimes
below 20~ns. For $2\ \mathrm{fb}^{-1}$, in run~II, the Tevatron should
significantly extend the sensitivity at large mass and lifetimes,
covering most of the lifetimes} for $m_{\tilde{G}}<1$~$\mathrm{keV}/c^2$
up to \textcolor{black}{a mass of around 150~}\munit.
The mass exclusion limit at 168~GeV for $\tau_{\tilde{\chi}}$~=~0~ns is comparable to the limit presented
in the D\O\ study of displaced photons in Ref.~\cite{key-36}, but for large lifetimes this result significantly extends the reach.

\subsection{Factors that might change the cross section limit}

\begin{figure}
\begin{center}\includegraphics[%
  width=8.6cm,
  keepaspectratio]{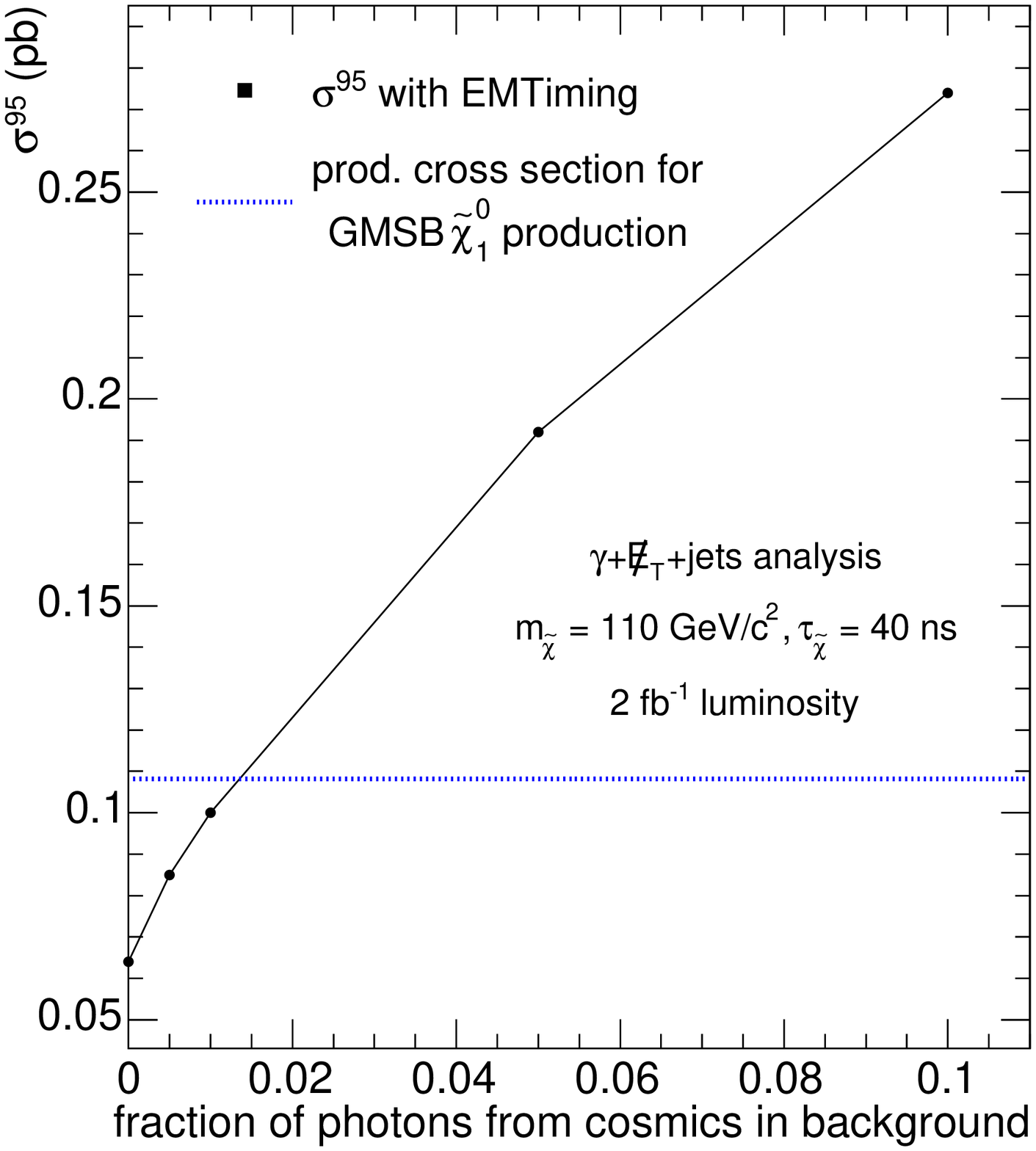}\end{center}

\caption{\label{cap:cosmicsfrac_gammaMETjets}The 95\%~C.L. cross section
limit on full GMSB production at \textcolor{black}{$m_{\tilde{\chi}}$~=~110~}\munit\ \textcolor{black}{ and}
$\tau_{\tilde{\chi}}$\textcolor{black}{~=~40~ns} as a function
of the fraction of the background from cosmic ray sources for a $\gamma$~+~$E_{T}\!\!\!\!\!\!\!/\,\,\,$~+~jets
analysis. The cross section limits rise approximately linearly as a function of the fraction and 10\% provides an outer bound on this fraction. A more reasonable fraction is probably 1-5\% which roughly doubles the limit.}
\end{figure}

\begin{figure}
\begin{center}\includegraphics[%
  width=8.6cm]{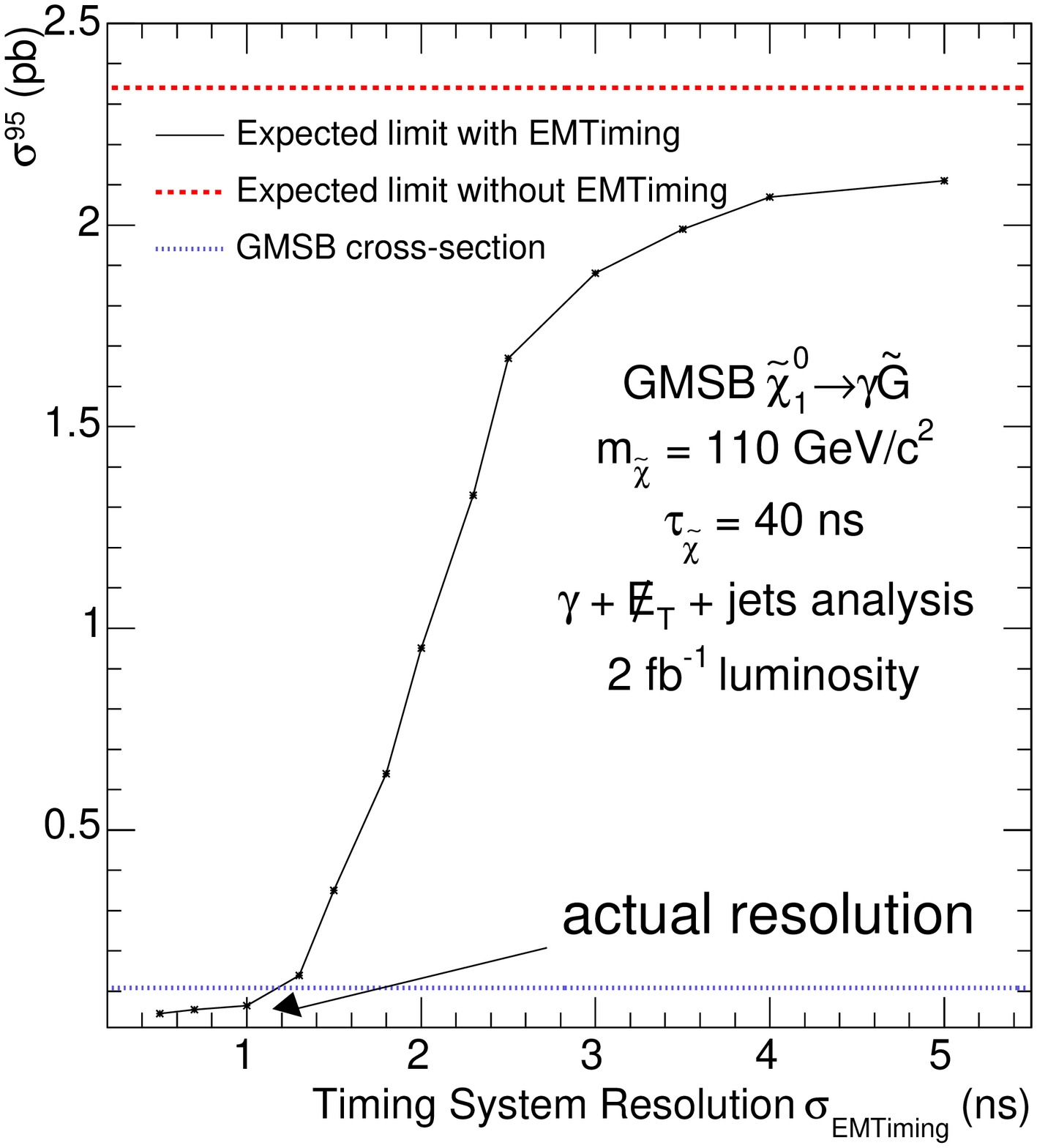}\end{center}

\caption{\textcolor{black}{\label{cap:TDCresolution}The 95\%~C.L. cross section limits vs. 
EMTiming system resolution for} \textcolor{red}{}\textcolor{black}{$m_{\tilde{\chi}}$~=~110~}\munit\
\textcolor{black}{and} $\tau_{\tilde{\chi}}$\textcolor{black}{~=~40~ns
at $2\ \mathrm{fb}^{-1}$ luminosity in the $\gamma$~+~}$E_{T}\!\!\!\!\!\!\!/\,\,\,$~+~jets
\textcolor{black}{analysis. As expected, for large resolution the
cross section with EMTiming approaches the cross section without EMTiming.
A system resolution of 1.0~ns improves the cross section limit 
by a factor of about 20, but this varies as a function of mass, lifetime and depends on the
analysis. It is reasonable to assume that the resolution will be within 20\% of the nominal 1.0~ns presented here.}}
\end{figure}

While we have taken the best available nominal values from the references for both the contamination of cosmic ray background events and the timing resolution in the $\gamma\gamma$~+~$E_{T}\!\!\!\!\!\!\!/\,\,\,$
and a $\gamma$~+~$E_{T}\!\!\!\!\!\!\!/\,\,\,$\textcolor{black}{~+~jets
analyses}, the limits are sensitive to mis-estimations of these values. For simplicity, rather than include them as a systematic error we estimate the variation of our results for these effects on our cross section limits for a neutralino mass of 110~\munit\
and a lifetime of 40~ns beyond the LEP~II exclusion region. Figure~\ref{cap:cosmicsfrac_gammaMETjets}
shows the cross section limit as a function of the fraction of events which are from cosmics
in the background sample. Using the same analysis style as in the
$\gamma$~+~$E_{T}\!\!\!\!\!\!\!/\,\,\,$~+~0~jets case we find
\textcolor{black}{cuts around $\Delta s_{1}=3.0$~ns} and $E_{T}\!\!\!\!\!\!\!/\,\,\,=55$~GeV,
\textcolor{black}{with $\Delta s_{2}$} varying from infinity down to
7~ns with a rising fraction of cosmics. The limits rise approximately linearly as a function of the fraction of events which are from cosmics. An upper bound on the fraction of cosmics of 
10\% would reduce the limits by about a factor of four; a more reasonable estimate is probably 1-5\% 
which would raise the limts by a factor of 2-3. 
The limits are potentially more sensitive to the resolution. Figure~\ref{cap:TDCresolution} 
shows how the limits change as a function of the system
resolution for the same mass and lifetime, for the original \textcolor{black}{$\gamma$~+}~$E_{T}\!\!\!\!\!\!\!/\,\,\,$~+~jets
analysis. While there is a mass/lifetime dependent resolution threshold from where the limit can change drastically and approach
the cross section without EMTiming, the limits are fairly stable (with the same factor of 2) for resolutions within 20\% of the nominal 1~ns resolution, and there is good reason to believe that the resolution will be better than advertised~\cite{key-26}.

\textcolor{black}{}%

\section{Conclusion}

We have studied the prospects of using the new EMTiming system at
CDF in the search for neutral, long-lived particles which
decay to photons, as one can find in supersymmetric models. We find
that the kinematic requirements and the EMTiming system provide excellent
rejection against SM backgrounds in complementary fashion. As the
mass increases the kinematics are more important and the sensitivity
gets better. For a given mass, as the lifetime increases more and
more of the neutralinos leave the detector and the overall sensitivity
goes down, but the EMTiming system provides additional rejection power
and allows for significant exclusions even at large lifetimes. While
the region where EMTiming produces the most additional rejection is
already excluded by LEP~II, the additional handle it provides should allow
the Tevatron in run~II to produce the world's most stringent limit
at masses above 80~\munit\ at high lifetimes and has the potential
to cover the entire \textcolor{black}{region} for $m_{\tilde{G}}<1$~$\mathrm{keV}/c^2$
up to \textcolor{black}{a neutralino mass of around 150~}\munit.

\begin{acknowledgments}
The authors would especially like to thank T.~Kamon for his early contributions to the
genesis of the idea of using the EMTiming system at CDF to search for GMSB
SUSY; it was his preliminary feasibility study which showed that the
analysis might be possible. We would also like to thank R. Arnowitt, R. Culbertson, B. Dutta, A. Garcia-Bellido, 
M. Goncharov, V. Khotilovich, V. Krutelyov, S.~W. Lee, P. McIntyre and S. Mrenna 
for help and useful discussions. We thank Fermilab and the FNAL
CAF system administrators for providing additional computing power.
This research was generously supported using funding from the College
of Science and the Department of Physics at Texas A\&M University. 
\end{acknowledgments}
\appendix

\section{Photon Pointing \label{sec:App_Photon_Pointing}}

\begin{figure}
\begin{center}\includegraphics[%
  width=8.6cm,
  keepaspectratio]{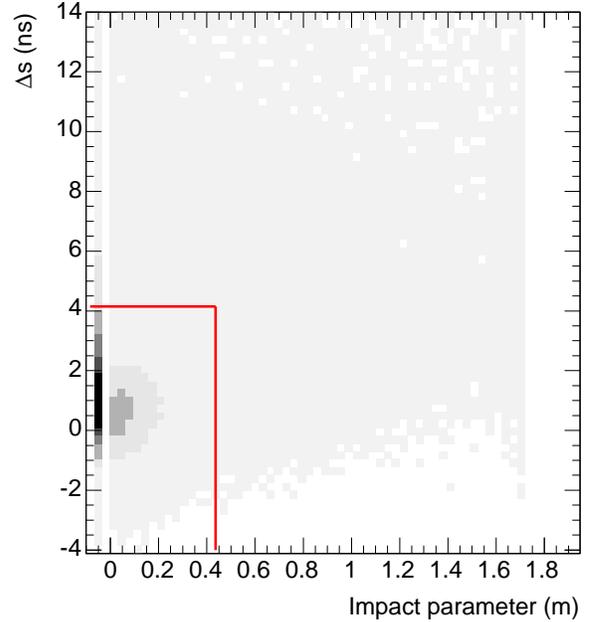}\end{center}

\caption{\label{cap:sig-b-deltas}The relationship between $\Delta s$ and
the impact parameter $b$ of a photon from \textcolor{black}{\small $\tilde{\chi}_{1}^{0}\rightarrow\gamma\tilde{G}$}
decays in a GMSB model with $m_{\tilde{\chi}}=70$~\munit\ and $\tau_{\tilde{\chi}}=10$~ns.
We show the selection requirements that give us the smallest 95\%~C.L. 
cross section limit in a $\gamma$~+~$E_{T}\!\!\!\!\!\!\!/\,\,\,$~+~jets
analysis. The photons without impact parameter measurement are assigned a $b=-0.1$~m. 
Due to the low cut on the impact parameter there are about as many
events in the low-$\Delta s$ high-$b$ as in the high-$\Delta s$
low-$b$ quarter, leading to a similar efficiency for a pure $b$-cut
compared to a pure $\Delta s$ cut. The combined restriction leads
to improved signal sensitivity.}
\end{figure}

\begin{figure}[H]
\begin{center}\includegraphics[%
  width=8.6cm,
  keepaspectratio]{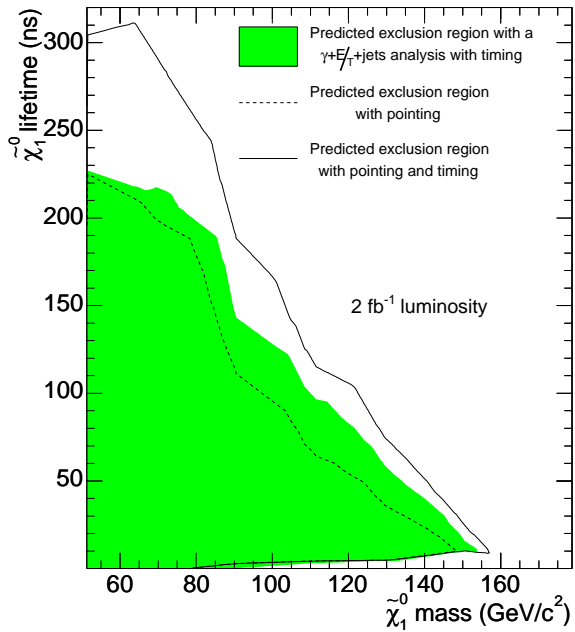}\end{center}

\caption{\label{cap:xsecmin-gmsb-2000lumi-pointing}\textcolor{black}{A comparison of the
expected} exclusion regions as a function
of neutralino mass and lifetime for the GMSB model at \textcolor{black}{$2\ \mathrm{fb}^{-1}$}
luminosity for a $\gamma$~+~$E_{T}\!\!\!\!\!\!\!/\,\,\,$~+~jets analysis with photon pointing and timing. While timing 
generally yields a higher sensitivity than pointing, both methods would, if we combined them, 
extend the exclusion region further than either of them alone. }
\end{figure}

As shown in Fig.~\textcolor{black}{\ref{cap:xsecmin-gmsb-2000lumi},
LEP~II has already excluded the low neutralino mass region using a} photon
pointing method~\cite{key-3}. In this section we compare the EMTiming system to
a potential photon pointing ability at CDF. A non-zero lifetime can
result in a macroscopic decay length and impact parameter, where the
impact parameter of the photon is basically the closest distance of
the trajectory to the collision point. While CDF has never
used its calorimeter for a pointing measurement it is possible to use the central EM strip/wire
gas chamber (CES) and the central pre-radiator gas chamber (CPR) at
CDF to measure two points along the photon trajectories that determine
the direction of the photon, and trace it back to yield the impact
parameter~\cite{key-14}. Since the CPR has no $z$-measurement ability
this allows only a measurement of the radial component of the impact
parameter with an estimated resolution of 10~cm (see Table \ref{cap:PointingErrors}).
One of the primary reasons this has not been used is the conversion, i.e. measurement, 
probability, of $\sim$65\%, with an angular
dependence obtained with: $$P_{C}=1-e^{-\frac{N_{\mathrm{rad}}}{\mathrm{sin}\theta}}\,\,\,,$$
where $N_{\mathrm{rad}}=1.072$ is the number of radiation lengths before the
CPR and $\theta$ is the angle with respect to the beamline. To estimate the sensitivity with a pointing method we consider
\textcolor{black}{a} $\gamma$~+~$E_{T}\!\!\!\!\!\!\!/\,\,\,$~+~jets
analysis. Figure~\ref{cap:sig-b-deltas} shows the distribution of
the signal events vs. impact parameter and $\Delta s$ in a $\gamma$~+~$E_{T}\!\!\!\!\!\!\!/\,\,\,$~+~jets
analysis taking into account the measurement probability; there are roughly as many events in the
region of low impact parameter and high $\Delta s$ as there are at
high impact parameter and low $\Delta s$. Hence either method should
have roughly the same effect on the exclusion region, as confirmed
by Fig.~\ref{cap:xsecmin-gmsb-2000lumi-pointing}, which shows the
expected exclusion region in the mass-lifetime plane. While timing is better than pointing by itself, 
if pointing turns out to be feasible, a combination of the two would improve the sensitivity.

Considered separately, a second advantage of timing is that it {}``filters''
manifestly long lifetime events, whereas the impact parameter allows
also short lifetime-high momentum events, which might come from SM. 
Another possible advantage of the combination is that in the event of an excess, we could draw
more information about the individual events, for instance determine the direction
of the photon. With the $x$-$y$-direction being fixed by the CPR/CES measurement,
we can use the timing system to measure the $z$-direction. Or if the
pointing provided the photon direction in $z$ and $x$-$y$, one could possibly
determine the position of the vertex and thus the decay time. However, with
the current 1.0~ns resolution the photon vertex position resolution
would be roughly 50~cm, if we assume the neutralino boost to be $\sim1.0$. 


%
\begin{table}

\caption{\textcolor{black}{\label{cap:PointingErrors}Photon pointing parameters
for the CDF detector \cite{key-14}. With this combination we estimate
an impact parameter resolution of 10~cm in the radial direction.}}

\begin{center}\textcolor{black}{}\begin{tabular}{lrp{8.6cm}}
\hline
\hline 
\multicolumn{2}{c}{measurement only in radial direction}\tabularnewline
Radius of CES&
184.15 cm\tabularnewline
Radius of CPR&
168.29 cm\tabularnewline
$\sigma_{\mathrm{CES}}$&
2~mm\tabularnewline
$\sigma_{\mathrm{CPR}}$&
5~mm\tabularnewline
$\mathrm{N_{rad}}$&
1.072 $\mathrm{X_{0}}$\tabularnewline
\hline
\hline
\end{tabular}\end{center}
\end{table}

\begin{thebibliography}{10}
\bibitem{key-25}CDF~II~Collaboration, F.~Abe \emph{et. al.,} FERMILAB-Pub-96/390-E.
\bibitem{key-24}S.~Kuhlmann \emph{et. al}., Nucl.~Instrum.~Methods Phys.~Res. A\textbf{~518},
39 (2004); CDF~Collaboration, P.~T.~Lukens, FERMILAB-TM-2198; for more details see http://hepr8.physics.tamu.edu/hep/emtiming/.
\bibitem{key-10}See for example CDF Collaboration, F.~Abe \emph{et. al}., Phys.~Rev. D\textbf{~59} 092002 (1999) 
and Phys.~Rev.~Lett. \textbf{81}, 1791 (1998).
\bibitem{key-29}Note that in principle the system can be used in conjunction with
the Time of Flight system to search for charged particles as well.
This becomes especially interesting for events in which the charged
particle decay in flight and no track is reconstructed.
\bibitem{key-2}See for example S. Ambrosanio \emph{et. al.},
Phys.~Rev. D\textbf{~54}, 5395 (1996); C.~H.~Chen and J.~F.~Gunion, Phys.~Rev. D\textbf{~58}, 075005 (1998).
\bibitem{key-30}\textcolor{black}{At CDF $x_{f}$ is measured by the shower-maximum detector (CES/PES)
in the electromagnetic calorimeter and $x_{i}$ is the collision point,
measured by the Silicon Vertex Chamber (SVX). The $t_{i}$ is calculated
by the Central Outer Tracker (COT) and the Time of Flight (TOF) system
which use the momentum and the measured time of flight of the charged
particles of the underlying event emerging from the collision point
and $t_{f}$ is the time of arrival of the photon from the EMTiming
system. }
\bibitem{key-26}M. Goncharov, priv. comm.
\bibitem{key-41}D\O~Collaboration, S. Abachi \emph{et. al.}, Nucl.~Instr.~Methods Phys.~Res. A\textbf{~338}, 185 (1994).
\bibitem{key-55}D\O\ Collaboration S.~Abachi  \emph{et. al.} Phys.~Rev.~Lett. \textbf{78}, 2070 (1997), and B.~Abbott \emph{et. al.} Phys.~Rev.~Lett. \textbf{80}, 442 (1998).

\bibitem{key-7}CDF~Collaboration, D. Acosta \emph{et. al.}, Phys.~Rev.~Lett. \textbf{89}, 281801 (2002).
\bibitem{key-8}D\O~Collaboration, B. Abbott \emph{et. al.}, Phys.~Rev.~Lett. \textbf{82}, 29 (1999).
\bibitem{key-39}Note that we do not assume that any analysis without EMTiming at CDF is robust enough to search for long-lived particles 
since there are no current collider experiments which have results without an additional handle such as photon pointing as 
done at LEP or D\O. For a discussion of photon pointing see for example ALEPH Collaboration, D. Decamp \emph{et. al.}, 
Nucl.~Instrum.~Methods Phys.~Res. A\textbf{~294}, 121 (1998) and D.~Cutts and G.~ Landsberg, arXiv:hep-ph/9904396.
\bibitem{key-3}ALEPH Collaboration, A. Heister \emph{et. al.}, Eur.~Phys.~J. C\textbf{~25},
339 (2002); A.~Garcia-Bellido, Ph.D. thesis, Royal Holloway University of London, 2002 (unpublished), arXiv:hep-ex/0212024.
\bibitem{key-27}See for example K. Olive, arXiv:hep-ph/9911307 and references therein. 
\bibitem{key-23} We follow B.~C.~Allanach \emph{et. al.}, Eur.~Phys.~J. C\textbf{25}, 113 (2002),
and take the messenger mass scale $M_{\mathrm{M}}$~=~2$\Lambda$, 
tan($\beta$)~=~15, sgn($\mu$)~=~1 and the number of messenger fields $N_{\mathrm{M}}$~=~1; 
$c_{\mathrm{Grav}}$ (gravitino mass factor) and $\Lambda$ (supersymmetry breaking scale) unrestricted.
\bibitem{key-5}T.~Sj\"ostrand, L.~L\"onnblad and S.~Mrenna, arXiv:hep-ph/0108264.
We used version 6.158.
\bibitem{key-56} F.~Paige and S.~Protopopescu, BNL Report BNL38034, 1986; F.~Paige, S.~Protopopescu, H.~Baer and X.~Tata, hep-ph/0001086. We use version 7.64 to generate the SUSY masses.
\bibitem{key-6}J. Conway, \textcolor{black}{http://www.physics.rutgers.edu/jconway/\linebreak
research/software/pgs/pgs.html. We used version~3.21.}
\bibitem{key-11}Note that we have chosen a neutralino mass of 70~\munit\  as an example, where the boost of the neutralino is lowest, 
to better illustrate the separation between single and diphoton case.
When we present cross section limits we move to a mass of 110~\munit\  
where timing provides sensitivity outside the current GMSB exclusion region from LEP~II.
\bibitem{key-21}E. Boos, A.~Vologdin, D.~Toback and J.~Gaspard, Phys.~Rev. D~\textbf{66},
013011 (2002).
\bibitem{key-31}Following Ref.~\cite{key-10} we have neglected the contribution from cosmic
ray backgrounds. The probability
of an event with a real photon and a fake photon from a cosmic
ray is small, and can be removed with simple topology cuts
with very high efficiency. Similar arguments hold for the even more unlikely case of two fake photons from cosmic ray background sources.
\bibitem{key-34}Following \cite{key-8} we have neglected the contribution from cosmic
ray backgrounds. The probability
of an event with two jets overlapping with a photon from a cosmic
ray is small, and and might be removed by requiring events where the photon is equal in magnitude 
and opposite in $\phi$ of the $E_{T}\!\!\!\!\!\!\!/\,\,\,$.
\bibitem{key-36}R. Culbertson \emph{et. al.}, hep-ph/0007080 (especially Fig. 18 therein).
\bibitem{key-14}L.~Balka \emph{et. al.}, Nucl.~Instr.~Methods Phys.~Res. A\textbf{~267}, 272 (1988); CDF Collaboration, F. Abe \emph{et. al.}, Phys.~Rev.~Lett.~\textbf{73}, 2662 (1994); 
D.~Partos, Ph.D. thesis, Brandeis University, 2001, FERMILAB-THESIS-2001-21.

\end{thebibliography}
\end{document}